\documentclass[11pt]{article}
\usepackage[margin=1in]{geometry}

\usepackage{microtype}

\usepackage{algorithm}
\usepackage{algorithmicx}
\usepackage[noend]{algpseudocode}
\usepackage[moderate]{savetrees}
\usepackage{enumitem}
\usepackage{amsthm,amsmath,amssymb,graphicx,mathrsfs,
subcaption}
\usepackage{array}
\usepackage{setspace}

\newcommand{\PK}[1]{}

\newcommand{\ceil}[1]{\lceil #1 \rceil}

\newcommand{\set}[1]{\{#1\}}
\newcommand{\ep}{\varepsilon}

\newcommand{\OPT}{\text{OPT}}

\newcommand{\dist}{d}
\newcommand{\cost}{\text{cost}}
\newcommand{\diam}{\text{diam}}
\newcommand{\SPC}{\text{SPC}}
\newcommand{\hdimension}{\eta}

\theoremstyle{plain}
\newtheorem{lemma}{Lemma}
\newtheorem{theorem}{Theorem}
\newtheorem*{restatelemma}{Lemma}

\newtheoremstyle{rstate}%
{}{}%
{\itshape}{}%
{\bfseries}{}
{ }{}

\theoremstyle{rstate}
\newtheorem*{restate}{Theorem}

\theoremstyle{definition}
\newtheorem{definition}{Definition}

\newcolumntype{M}[1]{>{\centering\arraybackslash}m{#1}}

\title{Polynomial-Time Approximation Schemes for $k$-Center and 
Bounded-Capacity Vehicle Routing in Graphs with Bounded Highway Dimension}

\author{Amariah Becker\thanks{Department of Computer Science, Brown
    University.  Email: \texttt{amariah\_becker@brown.edu}}\ $^{,\S}$, Philip
N. Klein\thanks{Department of Computer Science, Brown University.  Email:
\texttt{klein@brown.edu}}\ $^{,\S}$, David Saulpic\thanks{D\'{e}partement 
d'Informatique, \'{E}cole Normale Sup\'{e}rieure, 
Paris.  Work done while visiting Brown University.  Email:
\texttt{david.saulpic@ens.fr}}\ $^{,}$\thanks{
Research supported by National
Science Foundation grant CCF-1409520.}}
\date{}

\begin{document}

\maketitle

\begin{abstract} 
The concept of bounded highway dimension was developed to capture
observed properties of the metrics of road networks.  We show that a graph 
with bounded highway dimension, for any vertex,  can 
be embedded into a a graph of bounded treewidth in such a way that the distance 
between $u$ and $v$ is preserved up to an additive error of $\epsilon$ times the
distance from $u$ or $v$ to the selected vertex.  We show that this theorem 
yields a PTAS 
for {\sc Bounded-Capacity Vehicle Routing} in graphs of bounded highway 
dimension.  In this problem, the input specifies a depot and a set of clients, 
each
with a location and demand; the output is a set of depot-to-depot tours, where 
each
client is visited by some tour and each tour covers at most $Q$ units of client 
demand.
Our PTAS can be
extended to handle penalties for unvisited clients.

We extend this embedding result to handle a set $S$ of distinguished 
vertices. The treewidth depends on $|S|$, and the distance between $u$ and $v$ 
is preserved up to an additive error of $\epsilon$ times the
distance from $u$ and $v$ to $S$.

This embedding result implies a PTAS for {\sc Multiple Depot 
Bounded-Capacity Vehicle Routing}: the tours can go from one depot to another.
The embedding result also implies that, for fixed $k$, there is a PTAS for 
$k$-{\sc
  Center} in graphs of bounded highway dimension.  In this problem, the 
goal is to minimize $d$ such that there exist $k$ vertices (the {\em centers}) 
such that every vertex is within distance $d$ of some center.  Similarly,
for fixed $k$, there is a PTAS for $k$-{\sc Median} in graphs of
bounded highway dimension.  In this problem, the goal is to minimize
the sum of distances to the $k$ centers.
\end{abstract}

\setcounter{page}{0}
\newpage
\section{Introduction}\label{sec:intro}

The notion of \emph{highway dimension} was introduced by Abraham et 
al.~\cite{abraham10, abraham11} to explain the efficiency of some shortest-path 
heuristics. The motivation of this parameter comes from the work of Bast et al. 
\cite{bast06, bast07} who observed that, on a road network, a shortest path 
from a compact region to points that are far enough must go through one of a 
small number of nodes. They experimentally showed that the US 
road network has this property. Abraham et al.~\cite{abraham10, abraham11, 
abraham} proved results on the efficiency of shortest-path heuristics on 
graphs with bounded highway dimension.

For $r \in \mathbb{R}^+$ and $v \in V$, $B_v(r) = \{ u \in V | d(u, v) \leq
r\}$ denotes the ball with center $v$ and radius $r$. The definition of highway 
dimension
we use is from~\cite{feldmann}.  Let $c$ be a constant greater than 4.
\begin{definition} \label{def:hg_dim}
 The \emph{highway dimension} of a graph $G$ is the smallest integer 
$\hdimension$ 
such that for every $r \in \mathbb{R}^+$ 
and $v \in V$, there is a set of at most $\hdimension$ vertices in $B_{v}(cr)$ 
such that every shortest path in $B_{v}(cr)$ of length at least $r$ intersects 
this set.
\end{definition}
(See Section~\ref{sec:highway-dimension} for more on defining highway 
dimension.)

\subsection{New polynomial-time approximation schemes} \label{sec:new-schemes}

Abraham et al. note that ``conceivably, better algorithms for other
[optimization] problems can be developed and analyzed under the small
highway dimension assumption.''  Since some road networks are
described by graphs of small highway dimension, NP-hard optimization
problems that arise in road networks are natural candidates for study.
Feldmann~\cite{feldmannFPT} and Feldmann, Fung, K\"onemann, and 
Post~\cite{feldmann}
inaugurated this line of research, giving (respectively) a
constant-factor approximation algorithm for one problem and
quasi-polynomial-time approximation schemes for several other problems.
  In this paper, we give the first
\emph{polynomial-time approximation schemes} (PTASs) for classical
optimization problems in graphs of small highway dimension.

\subsubsection*{Vehicle routing}
Consider {\sc Capacitated Vehicle Routing}, defined as follows.  An
instance consists of a positive integer $Q$ (the \emph{capacity}), a
graph with edge-lengths, a subset $Z$ of vertices (called {\em
  clients}), a demand function $\rho:Z \rightarrow [1,2\ldots, Q]$,
and a distinguished vertex, called the {\em depot}.  A solution
consists of a set of \emph{tours}, where each tour is a path starts
and ends at the depot, and a function that assigns each client to a
tour that passes through it, such that the total client demand
assigned to each tour is at most $Q$.   (If a client $v$ is assigned
to a tour, we say that the tour \emph{visits} $v$.) The objective is
to minimize the sum of lengths of the tours.

We emphasize that in this version of {\sc Capacitated Vehicle Routing}, client 
demand
is \emph{indivisible}: a client's entire demand must be covered by a single 
tour.
For arbitrary metrics, the problem is APX-hard, even when $Q>0$ is 
fixed~\cite{asano}.
When $Q$ is unbounded, it is NP-hard to approximate to within a factor of~1.5 
even
when the metric is that of a star~\cite{golden}. Since stars have highway 
dimension
one, this hardness result holds for graphs of bounded highway
dimension.  We therefore require $Q$ to be constant.  To emphasize
this, we sometimes refer to the problem as {\sc Bounded-Capacity
  Vehicle Routing}.

\begin{theorem}\label{thm:cvr_PTAS}
For any $\epsilon>0$, $\hdimension > 0$ and $Q>0$,
  there is a polynomial-time algorithm that, given an instance
  of {\sc Bounded-Capacity Vehicle Routing} in which the capacity is $Q$
  and the graph has highway dimension $\hdimension$,
  finds a solution whose cost is at most $1+\epsilon$ times optimum.
\end{theorem}
Note that the running time is bounded by a polynomial whose degree
depends on $\ep$, $\hdimension$, and $Q$.  As we discuss in
Section~\ref{sec:related}, polynomial-time approximation schemes for
vehicle routing were previously known only for Euclidean spaces. (A
quasi-polynomial-time approximation scheme was known for planar
graphs.)

Our approach can be modified to handle a generalization in which an
instance also specifies a {\em penalty} for each client; the solution is 
allowed 
to omit some clients and the goal is to find a solution that minimizes the sum 
of costs plus penalties.  We call this {\sc Capacitated Vehicle Routing with 
Penalties}.

As we state in Theorem~\ref{thm:multiple-depot-vehicle-routing}, we
give a PTAS for a more general version of the problem, {\sc
  Multiple-Depot Bounded-Capacity Vehicle Routing}, in which there are a
constant number of depots, and each tour is required only to start and
end at one of the depots.  

\subsubsection*{$k$-{\sc Center} and $k$-{\sc Median}}
Next, consider $k$-{\sc Center} and $k$-{\sc Median} problems. 
Given a graph, the goal in $k$-{\sc Center} is to select a set of $k$ vertices 
(the
\emph{centers}) so as to minimize the maximum distance of a
vertex to the nearest center.  This problem might arise, for example,
in selecting locations for $k$ firehouses.  The objective in $k$-{\sc
  Median} is to minimize the average vertex-to-center distance.

For {\sc $k$-Center}, when the number $k$ of centers is unbounded, for any
$\delta>0$, it is NP-hard~\cite{gonzalez,hochbaum} to obtain a
$(2-\delta)$-approximation, even in the Euclidean plane under $L_1$ or
$L_\infty$ metrics,\footnote{Approximation better than 1.822 is hard
  under $L_2$, see~\cite{FederGreene}.}
even in unweighted
planar graphs~\cite{Plesnik}, and even in $n$-vertex graphs with
highway dimension $O(\log^2 n)$~\cite{feldmannFPT}.  It is not yet
known to be NP-hard in graphs with constant highway dimension, but
Feldmann~\cite{feldmannFPT} shows that, under the Exponential Time
Hypothesis, the running time of an algorithm achieving  a
$(2-\delta)$-approximation would be doubly exponential in a polynomial
in the highway dimension. 

These negative results suggest considering the problem in which $k$ is
bounded by a constant.  However, Feldmann~\cite{feldmannFPT} shows
that $(2-\epsilon)$-approximation is $W[2]$ hard for parameter $k$,
suggesting that the running time of any such approximation algorithm
would not be bounded by a polynomial whose degree is independent of
$k$.  Thus even for constant $k$, finding a much better approximation
seems to require that we restrict the metric.
Feldmann~\cite{feldmannFPT} gave a polynomial-time 3/2-approximation
algorithm for bounded-highway-dimension graphs, and raised the
question of whether a better approximation ratio could be achieved.
The following theorem answers that question.

\begin{theorem}\label{thm:kctr}  There is a function $f(\cdot, \cdot,
  \cdot)$ and a constant $c$ such that,
  for each of the problems {\sc $k$-Center} and {\sc $k$-Median},
  for any $\hdimension > 0, k > 0$
  and $\ep > 0$, there is an $f(\hdimension, k, \ep)n^{c}$ algorithm
  that, given an instance in which the graph has highway dimension at
  most $\hdimension$, finds a solution whose cost is at most $1+\ep$
  times optimum.
\end{theorem}
Note that the running time is bounded by a polynomial in $n$ whose
degree does {\em not} depend on $\hdimension$, $k$, or $\ep$.

\PK{Is the commented-out sentence compelling?}

\subsection{New metric embedding results}

The key to achieving the new approximation schemes is a new result on
metric embeddings of bounded-highway-dimension graphs into
bounded-treewidth graphs.  (\emph{Treewidth}, defined in
Section~\ref{sec:prelims}, is a measure of how complicated a graph is,
and many NP-hard optimization problems in graphs become
polynomial-time solvable when the input is restricted to graphs of
bounded treewidth.) 

A metric embedding of an (undirected) guest graph $G$ into a host graph $H$ 
(or, 
more
generally, metric space) is a mapping $\phi(\cdot)$ from the vertices
of $G$ to the vertices of $H$ such that, for every pair of vertices
$u,v$ in $G$, the $\phi(u)$-to-$\phi(v)$ distance in $H$ resembles the
$u$-to-$v$ distance in $G$.  Usually in studying metric embeddings one seeks
an embedding that preserves
$u$-to-$v$ distance up to some factor (the \emph{distortion}).  That is,
the allowed error is proportional to the original distance.  In this
work, the allowed error is instead proportional to the distance from a
given root vertex (or a constant number of vertices).

\begin{theorem}\label{thm:routing_embed}  There is a function
  $f(\cdot, \cdot)$ such that,
for every $\ep > 0$  graph $G$ of highway dimension $\hdimension$,
and vertex $s$, there exists a graph $H$ and an embedding $\phi(\cdot)$
of $G$ into $H$ such that 
\begin{itemize}[noitemsep]
 \item $H$ has treewidth at most $f(\ep,\hdimension)$, and
 \item for all vertices $u$ and $v$, 
$$\dist_G(u, v) \leq \dist_H(\phi(u), 
\phi(v)) \leq \dist_G(u, v) + \ep(\dist_G(s, u) + \dist_G(s, v))$$
\end{itemize}
\end{theorem}
We give a polynomial-time algorithm that, given the graph $G$,
constructs $H$ and the embedding of $G$ into $H$.

every

As we describe in greater detail in Section~\ref{sec:vehicle_routing}, our
PTAS for {\sc Bounded-Capacity Vehicle Routing} first applies
Theorem~\ref{thm:routing_embed} with $s$ being the depot and
$\epsilon'=\epsilon/c$ for a constant $c$ to be determined, obtaining
an embedding of the original graph into the bounded-treewidth graph
$H$.  The embedding induces an instance of {\sc Vehicle Routing} in
$H$.  The algorithm finds an optimal solution to this instance, and
converts it to a solution for the original instance.  This conversion
does not increase the cost of the solution.  However, we need to show
that the optimal solution in the original instance induces a solution
in $H$ of not too much greater cost.  We do this using a lower bound
due to Haimovich and Rinnoy Kan~\cite{haimovich}.



For the multiple-depot version of vehicle routing and for {\sc
  $k$-Center} and {\sc $k$-Median},
Theorem~\ref{thm:routing_embed} does not suffice.  We present a
generalization in which there is a set of root vertices, and the
allowed error is proportional to the minimum distance to any root
vertex.

\begin{theorem}\label{thm:multi}  There is a function
  $f(\cdot,\cdot,\cdot)$ such that,
 for every $\ep > 0$, graph $G$ of highway dimension $\hdimension$
and set $S$ of vertices of $G$, there exists a graph $H$ and an embedding 
$\phi(\cdot)$ of $G$ into $H$ such that 
\begin{itemize}[noitemsep]
 \item $H$ has treewidth $f(\hdimension, |S|, \ep)$, and
 \item for all vertices $u$ and $v$,
$$\dist_G(u, v) \leq \dist_H(\phi(u), \phi(v)) \leq 
(1+O(\ep))\dist_G(u, v) + \ep\min(\dist_G(S, u), \dist_G(S, v))$$
\end{itemize}
\end{theorem}

\subsection{Related Work} \label{sec:related}

\subsubsection*{Metric embeddings of bounded-highway-dimension graphs} 
  As mentioned in Section~\ref{sec:new-schemes},
Feldmann~\cite{feldmannFPT} and Feldmann et al.~\cite{feldmann}
inaugurated research into approximation algorithms for NP-hard
problems in bounded-highway-dimension graphs.  We discuss the work of
Feldmann~\cite{feldmann} soon.  Feldmann et al.~\cite{feldmann}
discovered quasi-polynomial-time approximation schemes for {\sc
  Traveling Salesman}, {\sc Steiner Tree}, and {\sc Facility
  Location}.  The key to their results is a probabilistic metric embedding of
bounded-highway dimension graphs into graphs of small treewidth.  The
\emph{aspect ratio} of a graph with edge-lengths is the ratio of the
maximum vertex-to-vertex distance to the minimum vertex-to-vertex
distance.  Feldmann et al. show that, for any $\epsilon>0$, for any
graph $G$ of highway dimension $\hdimension$, there is a probabilistic
embedding $\phi(\cdot)$ of $G$ of expected distortion $1+\epsilon$
into a randomly chosen graph $H$ whose
treewidth is polylogarithmic in the aspect ratio of $G$ (and also
depends on $\epsilon$ and $\hdimension$).  There are two obstacles to
using this embedding in achieving approximation schemes:
\begin{itemize}
\item The
distortion is achieved only in expectation.  That is, for each pair
$u,v$ of vertices, the expected $\phi(u)$-to-$\phi(v)$ distance in $H$ is at
most $(1+\epsilon)$ times the $u$-to-$v$ distance in $G$.
\item The treewidth depends on the aspect ratio of $G$, so is only
  bounded if the aspect ratio is bounded.
\end{itemize}
The first is an obstacle for problems (e.g. {\sc $k$-Center}) where
individual distances need to be bounded; this does not apply to
problems such as {\sc Traveling Salesman} or {\sc Vehicle Routing}
where the objective is a sum of lengths of paths.  The second is the
reason that Feldmann et al. obtain only quasi-polynomial-time
approximation schemes; it seems to be an obstacle to obtaining true
polynomial-time approximation schemes.

Nevertheless, the techniques introduced by Feldmann et al. are at the
core of our embedding results.  We build heavily on their framework.

\subsubsection*{Vehicle routing}
Haimovich and Rinnoy Kan~\cite{haimovich} proved the following
lower bound\footnote{Although their result addresses the unit-demand case, it 
generalizes
to instances where all non-zero client demand $\rho(v)$ is at least one for 
every client
$v \in Z$.}:

\begin{lemma}\label{lem:lw_bound}
 For {\sc Capacitated Vehicle Routing} with capacity $Q$, and client set $Z$,
$$\text{cost($\OPT$)} \geq \frac{2}{Q} \sum \set{d(c,s) \ :\ c\in Z}$$
\end{lemma}
Note that the {\sc Capacitated Vehicle Routing} problem, is a
generalization of {\sc Traveling Salesman} ($Q=n$, $Z=V$, and
$\rho(v)=1, \forall v$). Conversely, Haimovich and Rinnoy Kan show how
to use a solution to {\sc Traveling Salesman} to achieve a
constant-factor approximation for {\sc Capacitated Vehicle Routing},
where the constant depends on the approximation ratio for {\sc
  Traveling Salesman}.

Since {\sc Capacitated Vehicle Routing} in general graphs 
is APX-hard for every fixed $Q\geq 3$~\cite{asano1996covering,asano}, much 
work has focused on the Euclidean plane. Haimovich and Rinnoy 
Kan~\cite{haimovich} gave a polynomial-time approximation scheme (PTAS) for the 
Euclidean plane for the case when the capacity $Q$ is constant. Asano et 
al.~\cite{asano} showed how to improve this algorithm to get a PTAS when $Q$ is 
$O(\log n/\log \log n)$. For general capacities, Das and 
Mathieu~\cite{das-mathieu} gave a quasi-polynomial-time approximation
scheme for unbounded $Q$. 
Building on this work, Adamaszek, Czumaj, and Lingas~\cite{adamaszek} gave a 
PTAS that for any $\epsilon>0$ can handle $Q$ up to $2^{\log^\delta n}$ where 
$\delta$ depends on $\epsilon$.

Little is known for higher dimensions or other metrics. Kachay gave a PTAS 
in $R^d$ that requires $Q$ to be $O(\log^{1/d} \log n)$~\cite{Kachay2016}, 
and Hamaguchi and Katoh~\cite{hamaguchi} and Asano, Katoh, and 
Kawashima~\cite{AsanoKK} focused on constant-factor approximation algorithms 
for 
the case where the graph is a tree and client demand is divisible. Becker, 
Klein 
and
Saulpic~\cite{esa} gave
the first approximation scheme for a non-Euclidean metric: they describe a 
quasi-polynomial-time approximation scheme in planar graphs, but only
when the capacity $Q$ is polylogarithmic in the graph size. 
They introduce the idea of an error that depends on the distance to the 
depot, which we also use in the embedding presented in our work here.

\subsubsection*{$k$-{\sc Median}}
We have already surveyed some of the results on $k$-{\sc Center}.
Note that the results of Feldmann~\cite{feldmannFPT} are based on 
the definition of highway dimension of 2011~\cite{abraham11}, but can
be adapted to the definition we use here.

For {\sc $k$-Median}, constant-factor approximation algorithms have been
found for general metric spaces~\cite{shmoys1997approximation, jain-vazirani, 
arya}.
The best known approximation ratio for {\sc $k$-Median} in general metrics
 is $1 + \sqrt{3}$~\cite{li-svensson}, and it is NP-hard to approximate
 within a factor of $1+2/e$~\cite{guha-khuller}. For {\sc $k$-Median} in 
$d$-dimensional
 Euclidean space, PTAS have been found when
 $k$ is fixed (e.g.~\cite{bodoiu}) and when $d$ is fixed (e.g.~\cite{ARR98})
 but there
 exists no
 PTAS if $k$ and $d$ are part of the input \cite{guruswami}.  Recently 
Cohen-Addad
 et al.~\cite{Cohen-AddadKM16} gave a local search-based PTAS for {\sc 
$k$-Median} in edge-weight
 planar
 graphs that extends
 more generally to minor-closed graph families.

\bigskip
\subsection{Paper Outline}
Section~\ref{sec:prelims} provides preliminary definitions and presents useful 
results from Feldmann et al.~\cite{feldmann}. In Section~\ref{sec:embed} we 
give 
a first embedding result, that concerns graphs of bounded aspect ratio. 
Section~\ref{sec:main_embed} explains the
second embedding result (Theorem~\ref{thm:routing_embed}) and 
Section~\ref{sec:vehicle_routing}
presents how 
to use it to achieve a PTAS for {\sc Capacitated Vehicle Routing}, proving 
Theorem~\ref{thm:cvr_PTAS}. 
Section~\ref{sec:dp} describes the dynamic program used for {\sc Capacitated 
Vehicle Routing}, and finally Section~\ref{sec:multiple} gives the third 
embedding result (Theorem~\ref{thm:multi}) and applies it to several problems.

\section{Preliminaries}\label{sec:prelims}

We use $OPT$ to denote the optimum solution for an optimization 
problem. For minimization problems, an $\alpha$-approximation algorithm returns 
a solution with cost at most $\alpha\cdot \text{cost}(OPT)$.  
An \emph{approximation scheme} is a family of $(1+\ep)$-approximation 
algorithms indexed by $\ep >0$.  A \emph{polynomial-time approximation scheme} 
(PTAS) is an approximation scheme that for each $\ep$ runs in polynomial time.

For an undirected graph $G=(V,E)$, we use $d_G(u,v)$ (or $d(u,v)$ when $G$ is 
unambiguous) to denote the shortest-path distance between $u$ and $v$. For any 
vertex subsets $W\subseteq V$ and vertex $v \in V$ we let $d(v,W)$ denote 
$\min_{w\in W}{d(v,w)}$, and we let $diam(W)$ denote $\max_{u,v\in W}{d(u,v)}$.

An \emph{embedding} of a graph $G=(V,E)$ is a mapping $\phi$ from a 
\emph{guest} graph $G$ to a \emph{host} graph $H = (V,E_H)$. For notational 
simplicity, we identify the vertices of $H$ with points of $G$ and therefore 
omit $\phi$.

A \emph{tree decomposition} of a graph $G$ is a tree $T_G$ whose nodes are 
\emph{bags} of vertices that satisfy the following three criteria:
\begin{enumerate}[noitemsep]
\item Every $v \in V$ appears in at least one bag.
\item For every edge $(u,v) \in E$ there is some bag containing both $u$ and 
$v$.
\item For every $v \in V$, the bags containing $v$ form a connected subtree.
\end{enumerate}

The \emph{width} of $T_G$ is the size of the largest bag minus one.  The 
\emph{treewidth} of $G$ is the minimum width among all tree decompositions of 
$G$. It is a measure of how \emph{treelike} a graph is.  Observe that the 
treewidth of a tree is one.  Tree decompositions are useful for 
dynamic-programming
algorithms, but often give runtimes that are exponential in the 
treewidth.  Therefore many problems that are $NP$-hard in general can be solved 
efficiently in graphs of bounded treewidth.

Let $Y\subseteq X$ be a subset of elements in a metric space $(X,d)$. $Y$ is a 
\emph{$\delta$-covering} of $X$ if for all $x\in X$, $d(x,Y)\leq \delta$.  $Y$ 
is a 
\emph{$\beta$-packing} of $X$ if for all $y_1,y_2\in Y$ with $y_1 \neq y_2$, 
$d(y_1,y_2)\geq \beta$.  $Y$ is an \emph{$\ep$-net} if it is both an 
$\ep$-covering
and an $\ep$-packing.

\subsection{Shortest-Path Covers}

Instead of working directly with Definition~\ref{def:hg_dim}, we use the 
concept 
of
a \emph{shortest-path cover}, which as noted in Abraham et al. \cite{abraham11} 
is
a closely related and more convenient tool.

Recall that $c$ is a constant greater than 4.

\begin{definition}
 For a graph $G$ with vertex set $V$ and $r \in R^+$, a \emph{shortest-path 
cover} $\SPC(r) \subseteq V$ is a set of \emph{hubs} such that every shortest 
path of length in $(r,\ cr/2]$ contains at least one hub from this set. Such 
a cover is called \emph{locally s-sparse} for scale $r$ if every ball of 
diameter
$cr$ contains at most $s$ vertices from $\SPC(r)$.
\end{definition}

For a graph of highway dimension $\hdimension$, Abraham et al. \cite{abraham11} 
showed
how to find a $\hdimension \log \hdimension$-sparse shortest-path cover 
in polynomial time (though they show it for a different definition of highway 
dimension ($c = 4$), the algorithm can be straightforwardly adapted). This 
result justifies using shortest-path covers instead of 
directly using highway dimension.

\bigskip
\subsubsection{Town Decomposition}
Feldmann et al.\cite{feldmann} observed that a shortest-path cover for scale 
$r$ 
naturally
defines a clustering of the vertices into \emph{towns} \cite{feldmann}.
Informally, a \emph{town} at scale $r$ is a subset of vertices that are close 
to 
each
other and far from other towns and from the shortest-path cover for scale $r$. 
Formally,
a town is 
defined by at least one $v\in V$
such that $d(v,\SPC(r)) > 2r$ and is composed of $\{u\in V | d(u,v) \leq r\}$. 

\begin{figure}[!ht]
  \centering
  \includegraphics[width=.2\textwidth]{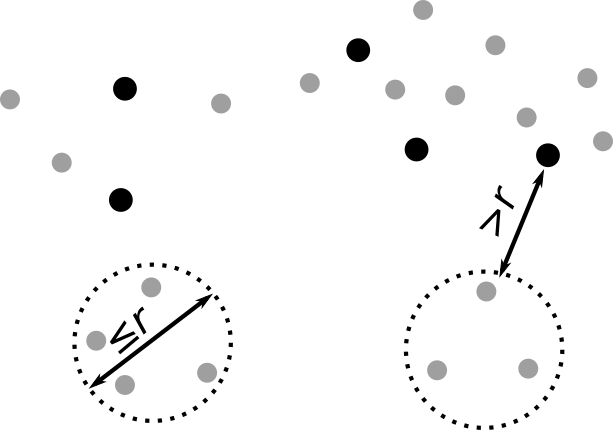}
  \caption{Illustration of Lemma~\ref{lem:towns}}
  \label{fig:towns}
\end{figure}

Lemma~\ref{lem:towns} describes key properties of towns proved in Feldmann et 
al and depicted in Figure~\ref{fig:towns}.

\begin{lemma}[Lemma 3.2 in \cite{feldmann}]\label{lem:towns}
If $T$ is a town at scale $r$, then
\begin{enumerate}[noitemsep]
\item $diam(T) \leq r$ and 
\item $d(T,V\setminus T) > r$ 
\end{enumerate}
\end{lemma}
\begin{proof}
 See Appendix~\ref{sec:town_proofs} for proof sketch.
\end{proof}

Feldmann et al. define a recursive decomposition of the graph using the concept 
of towns, which we adopt for this paper. First, scale all distances so that the 
shortest
point-to-point distance is a little more than $c/2$. Then fix a set of scales 
$r_i = (c/4)^i$.
We say that a town $T$ at scale $r_i$ is on \emph{level} $i$. 
Remark that the scaling ensures that $SPC(r_0) = \emptyset$, and therefore at 
level $0$ every vertex forms a singleton town.  Also note that the largest 
level 
is
$r_{max} = \ceil{\log_{c/4}diam(G_{scaled})} = \ceil{\log_{c/4}(\frac{c}
{2} \cdot \theta_G))}$, where $\theta_G$ is the
aspect
ratio of
the input graph.  Similarly at this topmost level, $SPC(r_{max}) = \emptyset$ 
since
there are no shortest paths that need to be covered. The only town at scale 
$r_{max}$
is the town that contains the entire graph.  We say that the town at scale 
$r_{max}$
and the singleton towns at scale $r_0$ are \emph{trivial} towns.  Since $c$ is 
a 
constant
greater than four, the total number of scales is linear in the input size.

Consider the set $\mathcal{T}=\{T \subseteq V | T\ \text{is a town on level}\ 
i\in \mathbb{N}\}$ of towns for all levels. Because of the properties of 
Lemma~\ref{lem:towns}, this set forms a laminar family and 
therefore has a tree structure. Towns on the same level are disjoint from each 
other.
By the isolation property of the town (the second property of 
Lemma~\ref{lem:towns}),
vertices outside of a town must be
far from the town, so a smaller town cannot be both inside and outside of a 
larger
town. Indeed, if two towns intersect, then the smaller town must be entirely 
inside
the bigger (see \cite{feldmann} for more details). Moreover, the decomposition 
has
the following properties:

\begin{lemma}[Lemma 3.3 in \cite{feldmann}]\label{lem:town_decomp}

For every town $T$ in a town decomposition $\mathcal{T}$,
\begin{enumerate}[noitemsep]
\item $T$ has either 0 children or at least 2 children, and 
\item if $T$ is a town at level $i$ and has child town $T'$ at level $j$, then 
$j<i$.
\end{enumerate}
\end{lemma}
\begin{proof}
 See Appendix~\ref{sec:town_proofs} for proof sketch.
\end{proof}

The set $\mathcal{T}$ is called the \emph{town decomposition} of G, with 
respect to the shortest-path cover, and is a key concept used in 
this paper. 

\subsubsection{Approximate Core Hubs}
Another insight that we adopt from Feldmann et al. is that rather than working 
with
\emph{all} hubs in the shortest-path covers, it is sufficient for approximation 
algorithms
to retain only a representative subset of the hubs: for $\ep > 0$, Feldmann et 
al.
define for each town $T$ a set $X_T$ of 
\emph{approximate core hubs} which is a subset of $T\ \bigcap\ \cup_i\SPC(r_i)$ 
with
the properties described in Lemma~\ref{lem:approx_core_hub}.
One key property of these sets is a bound on their \emph{doubling dimension}. 
The
doubling dimension of a metric is the smallest $\theta$ such that for every 
$r$, 
every
ball of radius $2r$ can be covered by at most $2^\theta$ balls of radius $r$.

\begin{lemma}[Theorem 4.2 and Lemma 5.1 in 
\cite{feldmann}]\label{lem:approx_core_hub}
Let $G$ be a graph of highway dimension $\hdimension$, and $\mathcal{T}$ be a 
town 
decomposition with respect to an inclusion-wise minimal, $s$-sparse 
shortest-path
cover. For any town
$T \in \mathcal{T}$,
\begin{enumerate}[noitemsep]
\item if $T_1$ and $T_2$ are different child towns of $T$, and $u \in T_1$ and 
$v \in T_2$, then there is some $h \in X_T$ such that $d(P[u,v],h) \leq \ep 
d(u,v)$, where $P[u,v]$ is the shortest $u$-to-$v$ path, and
\item the doubling dimension of $X_T$ is 
$\theta = O(\log(\hdimension s\log(1/\ep))$. 
\end{enumerate}
\end{lemma}

Intuitively, the shortest-path cover at scale $r_i$ forms a set of points that 
covers \textit{exactly} the shortest-paths of length in $(r_i, r_{i+1}]$. 
Therefore to cover the shortest-path between $u \in T_1$ and $v \in T_2$, we 
can use a hub at level $j$ such that $d(u, v) \in (r_j, r_{j+1}]$. 
Unfortunately, taking all the shortest-path covers gives a set too big for our 
purpose. But since we want to cover \textit{approximately} the distances 
between points in different child towns, we can take only a subset of the 
shortest-path covers that has a low doubling dimension. This subset can be 
found 
in
$n^{O(1)}$ time as explained in Feldmann et al.~\cite{feldmann}.

\subsubsection{Minimality of Shortest-Path Covers}
Note that the result of Lemma~\ref{lem:approx_core_hub} requires the 
shortest-path
covers be inclusion-wise minimal.  For the embedding we present in 
Section~\ref{sec:main_embed},
however, it is useful to assume
that the depot is not a member of any town except for the trivial topmost
town containing
all of $G$ and bottommost singleton town containing just the depot.  One way to 
ensure
this is to add the depot to the shortest-path cover at every level, but this 
violates
the minimality requirement (see Appendix~\ref{sec:add_depot} for further 
discussion).
Lemma~\ref{lem:depot_safety}, however,
shows that this assumption can be made \emph{safely} without asymptotically 
changing
our results.

We prove a more general statement for a set $S$ of depots. Recall that $c$ is a 
constant
greater than four and that all edges have been scaled
so that the smallest point-to-point distance is slightly more than $c/2$.

\begin{lemma}\label{lem:depot_safety}
Any graph $G = (V,E)$ with highway dimension $\hdimension$,
diameter $\Delta_G$, and designated vertex set  $S \subseteq V$ can be modified 
by
adding
$O(\hdimension^2|S|^3\log\Delta_G)$ new vertices and edges, such
that the resulting graph $G'=(V',E')$ 
\begin{itemize}
\setlength\itemsep{0em}
\item has highway dimension at most $\hdimension+|S|$ 
\item for all $u,v, \in V'$, $d_{G'}(u,v) \in (\frac{c} {2},\ \frac{3c} 
{4}\Delta_G]$
\item for all $u,v \in V$, $d_{G'}(u,v) = d_G(u,v)$, and 
\item for every $s \in S$, the only towns containing $s$ in the town 
decomposition
of $G'$ are the trivial towns.
\end{itemize}
\end{lemma}

\begin{proof}
See Appendix~\ref{sec:add_depot}.
\end{proof}

Note that after applying the modification of Lemma~\ref{lem:depot_safety} to the
(scaled) input graph, the resulting graph has size that is polynomial in the 
size
of the original input.

\section{Embedding for Graphs of Bounded Diameter}\label{sec:embed}

Lemma~\ref{thm:embed} describes an embedding for the case when the graph has 
bounded
diameter.
This embedding gives only a small \emph{additive} error, and will prove to be a 
useful tool for the following sections. In this section we show how to 
construct 
this embedding. 

\begin{lemma}\label{thm:embed}
There is a function $f(x,y)$ such that, for any $\ep > 0$ and $\hdimension>0$, 
for 
any graph $G$ with highway dimension at most $\hdimension$ and diameter
$\Delta$,
there is a graph $H$ with treewidth at most $f(\ep,\hdimension)$ and an 
embedding 
$\phi(\cdot)$ of $G$ into $H$ such that, for all points $u$ and $v$,
$$\dist_G(u, v) \leq \dist_H(\phi(u), \phi(v)) \leq \dist_G(u, v) + 4\ep\Delta$$
Furthermore, there is a polynomial-time algorithm to construct $H$ and the 
embedding.
\end{lemma}

\begin{figure}[!ht]
\centering
    \begin{subfigure}[t]{0.32\textwidth}
        \includegraphics[width=\textwidth]{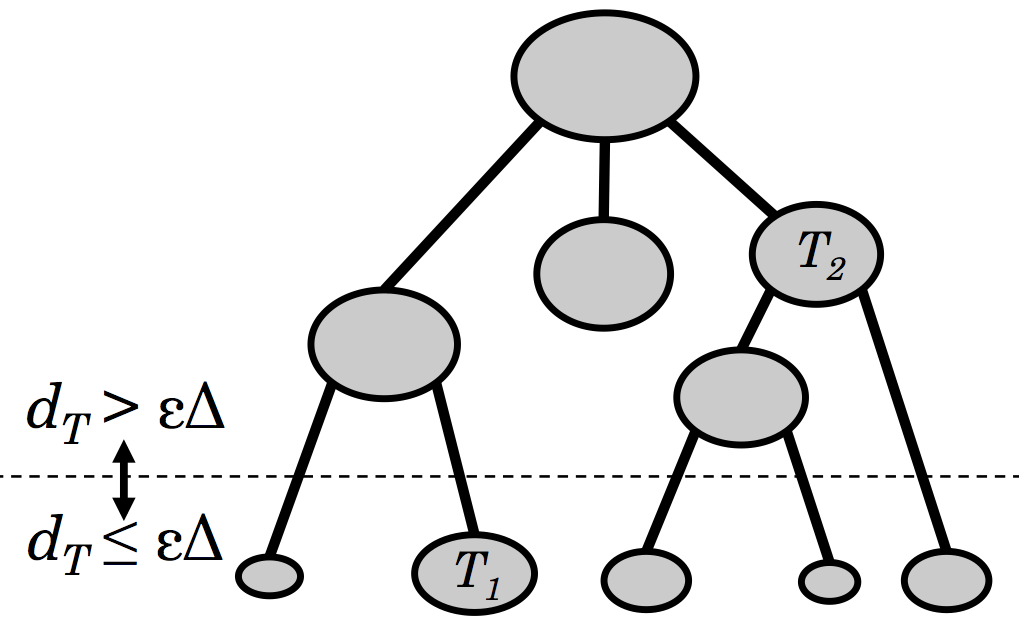}
        \caption{Town decomposition}
        \label{fig:const_diam_decomp}
    \end{subfigure}
    \quad\vline\quad
    \begin{subfigure}[t]{0.12\textwidth}
        \includegraphics[width=\textwidth]{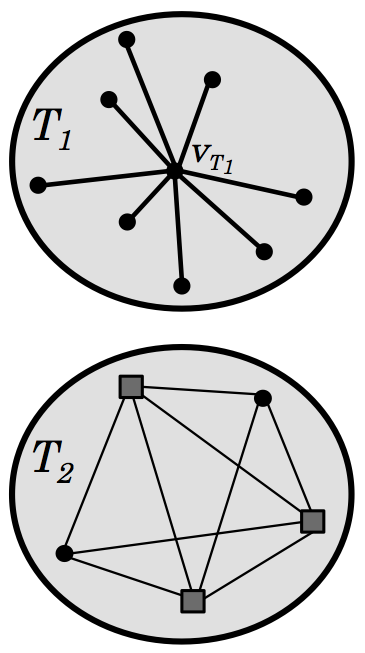}
        \caption{Embeddings}
        \label{fig:const_diam_embed}
    \end{subfigure}
    \quad\vline\quad
    \begin{subfigure}[t]{0.18\textwidth}
        \includegraphics[width=\textwidth]{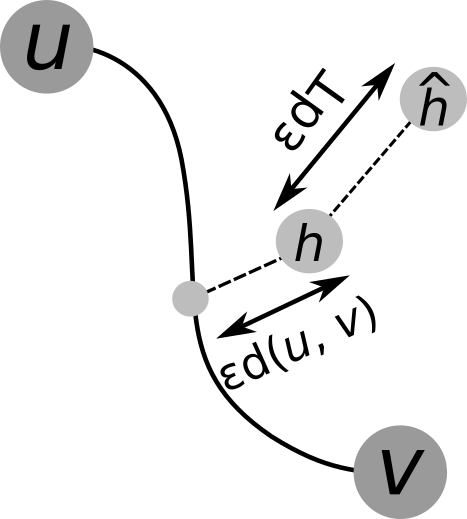}
        \caption{Path approximation}
        \label{fig:approx}
    \end{subfigure}
    \caption{\label{fig:const_diam}(a) An example town decomposition. $T_1$ has 
diameter at most $\ep\Delta$ and $T_2$ has diameter greater than $\ep\Delta$.
    (b) Two cases of town embeddings. $T_1$ is embedded as a star with center 
$v_
    {T_1}$. The embedding of $T_2$ connects all vertices in $T_2$ to all hubs 
in 
$
    \hat{X}_{T_2}$(depicted as squares). (c) Hub $\hat{h} \in \hat{X}_T$ is 
close
    to hub $h \in X_T$ which itself is close to the shortest $u$-to-$v$ path.}  
 
\end{figure}

We first present an algorithm to compute the host graph $H$ and a tree 
decomposition of $H$. This algorithm relies on the town decomposition 
$\mathcal{T}$ of $G$, described in Section~\ref{sec:prelims}.

The host graph $H$ is constructed as follows. First, consider a town $T$ that 
has diameter $d \leq \ep \Delta$ but has no ancestor towns of diameter $\ep 
\Delta$ or smaller. We call such a town a \emph{maximal} town of diameter at 
most $\ep \Delta$. The town $T$ is embedded into a star: choose an arbitrary
vertex $v_T$ in $T$, and for each $u \in T$, include an edge in $H$ between 
$u$ and $v_T$ with length $d_G(u,v_T)$ equal to their distance in $G$ (see 
Figures~\ref{fig:const_diam_decomp} and \ref{fig:const_diam_embed}).

Now consider a town $T$ of diameter $d_T > \ep \Delta$. The set of approximate
core hubs $X_T$ can be used as \emph{portals} to preserve distances between 
vertices lying in different child towns of $T$.  Specifically, by 
Lemma~\ref{lem:approx_core_hub}, for every pair of vertices $(u,v)$ in 
different 
child towns of $T$, $X_T$ contains a vertex that is close to the shortest path 
between $u$ and $v$. In order to approximate the shortest paths,  it is 
therefore sufficient to consider a set of points \emph{close to} $X_T$. Let 
$\hat{X}_T$ be an $\ep d_T$-net of $X_T$. For each $\hat h \in \hat{X}_T$ and 
$v\in T$, include an edge in $H$ connecting $v$ to $\hat h$ with length 
$d_H(v,\hat h)=d_G(v,\hat h)$ equal to the $v$-to-$\hat h$ distance in $G$ (see 
Figures~\ref{fig:const_diam_decomp} and \ref{fig:const_diam_embed}).

\bigskip
The tree decomposition $D$ mimics the town decomposition tree: for each town 
$T$ of diameter greater than $\ep \Delta$, there is a bag $b_T$. This bag is 
connected in $D$ to all of the bags of child towns of $T$ and contains all of 
the vertices of the net assigned to $T$ and of the nets assigned to $T$'s 
ancestors in the town decomposition. Formally, if $A_T$ denotes the set of all 
towns that contain $T$, $b_T = \bigcup_{T' \in A_T} \hat{X}_{T'}$.  Note that 
if 
$T'$ is the parent of $T$ in the town decomposition, $b_T = \hat{X}_T \cup 
b_{T'}$. Now for each maximal town $T$ of diameter at most $\ep \Delta$ with 
parent town $T'$, the tree decomposition contains a bag $b_T^0$ connected to a 
bag $b_T^u$ for each vertex $u \in T$. We define $b_T^0 = \{v_T\} \cup b_{T'}$ 
and $b_T^u = \{u\} \cup b_T^0$.

Following Feldmann et al. \cite{feldmann}, the above construction can be shown 
to be polynomial-time constructible. The following three lemmas therefore prove 
Lemma~\ref{thm:embed}.

\begin{lemma}
$D$ is a valid tree decomposition of $H$.
\end{lemma}

\begin{proof} For $D$ to be a valid tree decomposition of $H$, it has to satisfy
the three criteria listed in the preliminaries.

As every vertex $v$ is in some maximal town $T$ of diameter at most 
$\ep\Delta$ (because every vertex form a singleton town at level $0$), there is 
a leaf $b_T^v$ of $D$ that contains $v$. Moreover, this leaf contains all of 
the 
vertices adjacent to $v$ in $H$: if an edge connects $u$ and $v$, then either 
$u$ or $v$ is the center of the star for $T$, or $u$ is in the net of some town 
that contains $v$. In both cases the construction of $D$ ensure that $u$ is in 
$b_T^v$. Finally, let $T$ be a town such that $b_T$ is the \emph{highest} bag 
in 
the tree decomposition that contains $v$. As the towns at a given height of the 
town decomposition form a partition of the vertices, this town is unique. Since 
the town decomposition has a laminar structure, $v$ cannot appear in a bag that 
is not a descendant of $b_T$. Furthermore, by definition of the bags, $v$ 
appears in all descendants of $b_T$, proving the third property.
\end{proof} 

\begin{lemma}\label{lem:tw}
$H$ has a treewidth $O((\frac{1}{\ep})^\theta \log_{\frac{c}{4}} 
\frac{1}{\ep})$, where $\theta$ is a bound on the doubling dimension of the 
sets $X_T$.
\end{lemma}

\begin{proof}
Since the size of the bags is clearly bounded by the depth times the maximal 
cardinality of $\hat{X}_T$, it is enough to prove that, for each town $T$, 
$\hat{X}_T$ is bounded by $(\frac{1}{\ep})^\theta$, and that the tree 
decomposition has a depth $O(\log_{\frac{c}{4}} \frac{1}{\ep})$.

By Lemma~\ref{lem:approx_core_hub}, the doubling dimension of $X_T$ is bounded 
by $\theta$. $\hat{X}_T$ is a subset of $X_T$, so its doubling dimension is 
bounded by $2\theta$ (see Gupta et al. \cite{GKL03}).
Furthermore, the aspect ratio of $\hat{X}_T$ is $\frac{1}{\ep}$: the longest 
distance between members of $\hat{X}_T$ is bounded by the diameter $d_T$ of the
town, and the smallest distance is at least $\ep d_T$ by definition of a net. 
The cardinality of a set with doubling dimension $x$ and aspect ratio $\gamma$ 
is bounded by $2^{x \lceil \log_2 \gamma \rceil}$ (see \cite{gupta} for a 
proof), therefore $|\hat{X}_T|$ is bounded by $(\frac{1}{\ep})^\theta$.

We prove now that the tree decomposition has a depth $O(\log_{\frac{c}{4}} 
\frac{1}{\ep})$. Let $T$ be a town of diameter $d_T > \ep \Delta$ and let $r_i$ 
be the scale of that town. By Lemma~\ref{lem:towns}, $d_T \leq r_i$, and since 
$r_i=(\frac{c}{4})^i$ and $d_T > \ep \Delta$, we can conclude that $i > 
\log_{\frac{c}{4}} \ep \Delta$. As the diameter of the graph is  $\Delta$, the 
biggest town has a diameter at most $\Delta$. It follows that $r_i \leq \Delta$ 
and therefore $i \leq \log_{\frac{c}{4}} \Delta$. The depth of $b_T$ in the 
tree 
decomposition is therefore bounded by $\log_{\frac{c}{4}} \frac{\Delta}{\ep 
\Delta} = \log_{\frac{c}{4}} \frac{1}{\ep}$. Furthermore, the tree 
decomposition 
of a town of diameter at most $\ep \Delta$ has depth $2$. The overall depth is 
therefore $O(\log_{\frac{c}{4}}\frac{1}{\ep})$, concluding the proof.
\end{proof}

\begin{lemma}\label{lem:stretch}
For all vertices $u$ and $v$, $\dist_G(u, v) \leq \dist_H(u, v) \leq \dist_G(u, 
v) + 4\ep\Delta$
\end{lemma}

\begin{proof}
Let $u$ and $v$ be vertices in $V$, and let $T$ be the town that contains both 
$u$ and $v$ such that $u$ and $v$ are in different child towns of $T$.

If $T$ has diameter $d_T \leq \ep \Delta$, then let $T'$ be the maximal town of 
diameter at most $\ep \Delta$ that is an ancestor of $T$ (possibly $T$ itself). 
By construction, $T'$ was embedded into a star centered at some vertex $v_{T'} 
\in T'$, so $\dist_H(u, v) \leq \dist_H(u, v_{T'}) + \dist_H(v_{T'}, v) \leq 
\dist_G(u, v_{T'}) + \dist_G(v_{T'}, v) \leq 2 \ep \Delta$.

Otherwise if $T$ has diameter $d_T > \ep \Delta$, then by 
Lemma~\ref{lem:approx_core_hub}, there is some $h \in X_T$ such that 
$\dist_G(P[u,v],h) \leq \ep \dist(u,v)$.  Since $\hat{X}_T$ is an $\ep d_T$ 
cover of $X_T$, there is some $\hat{h} \in \hat{X}_T$ such that 
$\dist(h,\hat{h})\leq \ep d_T$. The host graph $H$ includes edges $(u,\hat{h})$ 
and $(\hat{h},v)$, so 

$\dist_H(u,v) \leq \dist_H(u, \hat{h}) + \dist_H(\hat{h}, 
v) \leq \dist_G(u,h) + \dist_G(h,v) + 2\ep \dist(u,v) + 2\ep d_T \leq 
\dist_G(u,v) +4 \ep \Delta$ (see Figure~\ref{fig:approx}).

\noindent Finally, since edge lengths in $H$ are given by distances in $G$, 
$\dist_G
(u, v)
\leq \dist_H(u, v)$ for all $u,v\in V$.
\end{proof}

The next sections present some applications of the above 
embedding.

\section{Main Embedding}\label{sec:main_embed}
In this section, we prove Theorem~\ref{thm:routing_embed}, restated here for 
convenience.

\begin{restate}{\bf \ref{thm:routing_embed}.}
There is a function
  $f(\cdot, \cdot)$ such that,
for every $\hat \epsilon > 0$  graph $G$ of highway dimension $\hdimension$,
and vertex $s$, there exists a graph $H$ and an embedding $\phi(\cdot)$
of $G$ into $H$ such that 
\begin{itemize}[noitemsep]
 \item $H$ has treewidth at most $f(\hat \epsilon,\hdimension)$, and
 \item for all vertices $u$ and $v$, 
$$\dist_G(u, v) \leq \dist_H(\phi(u), 
\phi(v)) \leq \dist_G(u, v) + \hat \epsilon(\dist_G(s, u) + \dist_G(s, v))$$
\end{itemize}
\end{restate}

\bigskip
\subsection{Embedding Construction}\label{sec:embed_construction}
Given the parameter $\hat \epsilon$, our goal for the embedding is that
$$\dist_G(u, v) \leq \dist_H(\phi(u), \phi(v)) \leq 
\dist_G(u, v) + \hat \ep(\dist_G(s, u) + \dist_G(s, v))$$
With this goal in mind, we define $\epsilon = \min \set{1/4, \hat \ep/c}$ for
an appropriate constant $c$, and we prove that
$$\dist_G(u, v) \leq \dist_H(\phi(u), \phi(v)) \leq 
\dist_G(u, v) + O(\ep)(\dist_G(s, u) + \dist_G(s, v))$$
The constant $c$ is chosen to compensate for the big-O in the above inequality.

Our construction relies on the assumption that the
depot $s$ does not appear in any non-trivial town. By 
Lemma~\ref{lem:depot_safety}
(using
$S = \{s\}$), this assumption is
\emph{safe}, since the input graph can be modified to satisfy this assumption 
without
(asymptotically) changing the highway dimension, diameter, or size of the 
graph. 
Furthermore since the modification preserves original distances, and all newly 
added
vertices can be assumed to have no client demand, the modification does not 
affect
the solution.

The root town in the composition, denoted $T_0$, is the town that contains the 
entire
graph. We say that a town $T$ that is a child of the root town is a 
\emph{top-level
town}, which means that the only town that properly contains $T$ is $T_0$. 

The assumption that the depot, $s$, does not appear in any non-trivial town 
implies
that the top-level town that contains $s$ is the trivial singleton town. This 
assumption
is helpful to bound the 
distance between a top-level town $T$ and the depot $s$: as $s \notin T$, 
Lemma~\ref{lem:towns}
gives the bound $\dist(T, s) \geq \diam(T)$. This bound turns out to be very 
helpful
in the construction of the host graph.

We use Lemma~\ref{thm:embed} to construct an embedding for each top-level 
town. It remains to connect these embeddings : we cannot approximate $X_{T_0}$ 
with a net as we did in Lemma~\ref{thm:embed}, because the diameter of $G$ 
may be arbitrarily large.

To cope with that issue, we define inductively the hub sets $X_0^0, X_0^1, ...$ 
such that $X_0^k$ is a net of $X_{T_0} \cap B_s(2^k)$. Let $X_0^0$ be an 
$\ep$-net of $X_{T_0}\cap B_s(1)$ that contains the depot, $s$, and for $k \geq 
0$ let $X_0^{k+1}$ be an $\ep 2^{k+1}$-net of the set $\big(X_{T_0} \cap 
(B_s(2^{k+1}) - B_s(2^k))\big)\ \bigcup \ X_0^k$ that contains the depot. This 
construction ensures that $X_0^{k+1} \cap B_s(2^{k}) \subseteq X_0^{k}$, which 
will be helpful in Section~\ref{ssec:treedec} to find a tree decomposition of 
the host graph. Note that we can assume $s \in X_{T_0}$, since adding it 
increases
the doubling dimension by at most one and thus does not
change the result of Lemma~\ref{lem:approx_core_hub}.

\begin{figure}[!h]
\centering
\caption{\label{fig:routing_embed} (a) Towns $T_1$ and $T_2$ are top-level
towns, with $l(T_1)= i$ and $l(T_2) = i+1$.
(b) The embedding of each top-level town (shown as circles) are connected to a 
band
of $\log_2 \frac{1}{\ep} + 1$ hub sets (shown as squares). Edges are striped to
convey
that they connect \emph{all} vertices of the given hub-set endpoint to 
\emph{all}
vertices of the town-embedding endpoint.
(c) The vertices of each bag $\mathcal{B}$ (shown as circles) are added to each 
bag
of each descendent
top-level-town tree decomposition (shown as triangles).}

    \begin{subfigure}[t]{0.3\textwidth}
      \includegraphics[width=\textwidth]{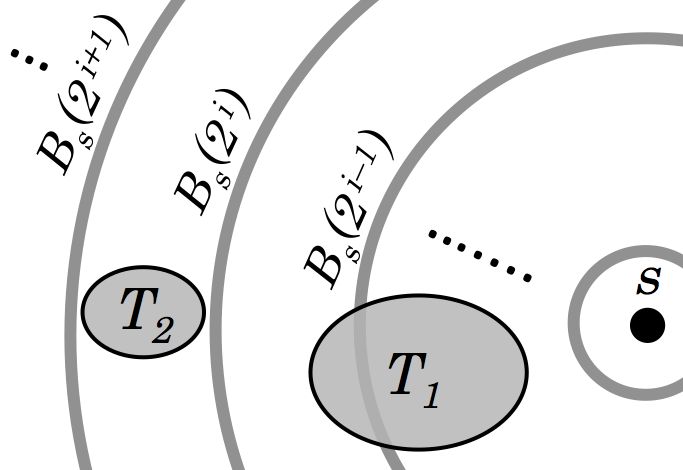}
      \caption{Towns}
      \label{fig:town_levels}
    \end{subfigure}
    \quad\vline\quad
    \begin{subfigure}[t]{0.45\textwidth}
      \includegraphics[width=\textwidth]{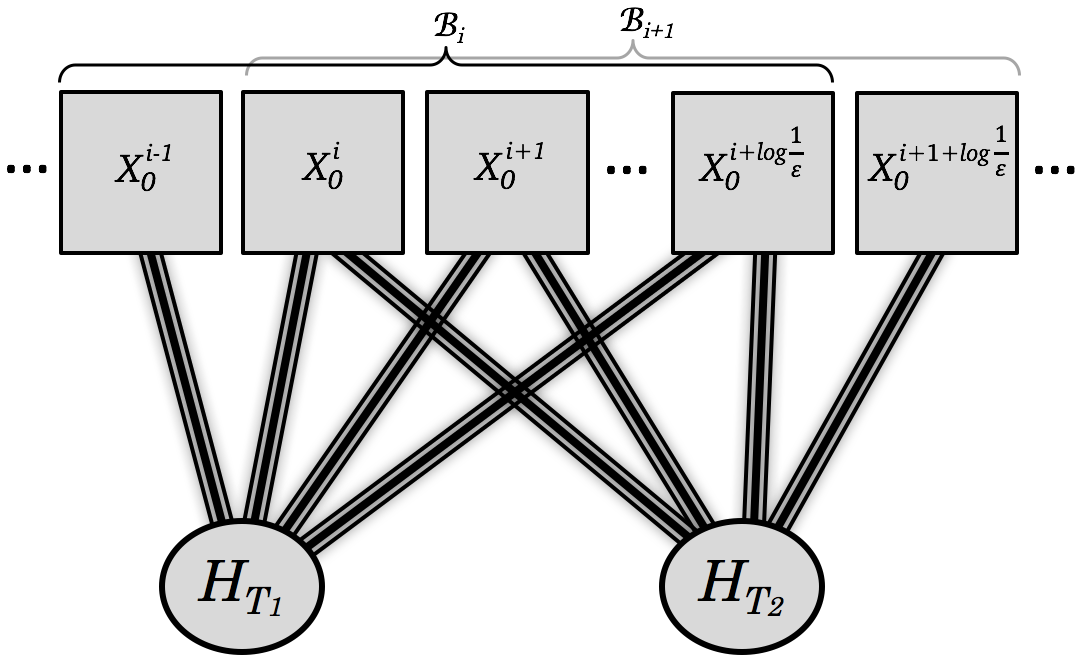}
      \caption{Embedding}
      \label{fig:town_hub_embed}
    \end{subfigure}\\

    \hrulefill
    
    \begin{subfigure}[t]{0.25\textwidth}
      \includegraphics[width=\textwidth]{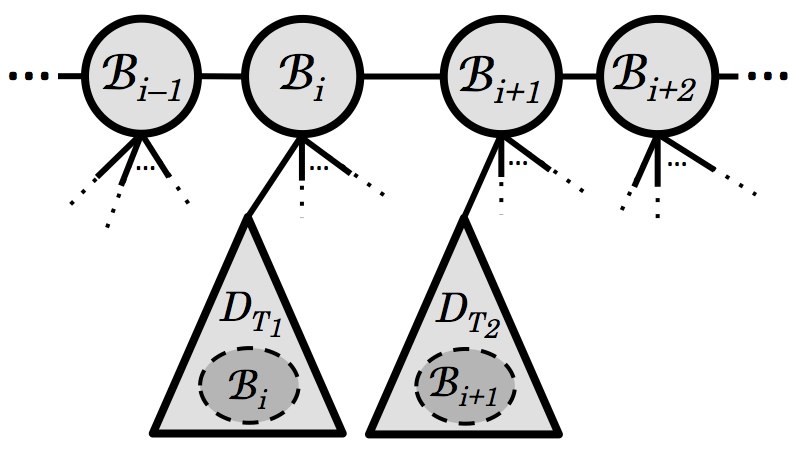}
      \caption{Tree decomposition}
      \label{fig:tree_decomp}
    \end{subfigure}
\end{figure}

For a set of vertices $\mathcal{X}\subseteq V$, we define 
$l(\mathcal{X}) = \lceil \log_2 ( \max_{v \in \mathcal{X}} \dist(s, v) ) 
\rceil$ 
(See Figure~\ref{fig:town_levels}).

For every child town $T$ of $T_0$, the host graph connects every vertex $v$ of 
$T$
to every hub $h$ in $X_0^{l(T)}, \ldots, X_0^{l(T) + \log_2(1/ \ep)}$ with an 
edge
of length $d_G(v,h)$ (See Figure~\ref{fig:town_hub_embed}).

\bigskip

\bigskip
\subsection{Proof of Error Bound} 

In Lemma~\ref{lem:embed_error} we prove a bound on the error incurred by the 
embedding.  Our proof makes use of the following lemma.

\begin{lemma}\label{lem:hub_cover}
 For all $k$, $X_0^{k}$ is an $\ep2^{k+1}$-covering of $X_{T_0} \cap B_s(2^k)$.
\end{lemma}

\begin{proof}
We proceed by induction. By construction, $X_0^0$ is an $\ep$-net (and thus 
also 
an
$\ep2^{1}$-covering)
of $X_
{T_0}\cap
B_s(2^0)$.  Assume that $X_0^{k-1}$ is an $\ep2^{k}$-covering of $X_{T_0} \cap 
B_s
(2^{k-1})$, and
let $x \in X_{T_0} \cap B_s(2^k)$.

$X_0^{k}$ is an $\ep 2^{k}$-net of the set $\big(X_{T_0} \cap 
(B_s(2^{k}) - B_s(2^{k-1}))\big)\ \bigcup \ X_0^k$, so if $x \in B_s(2^{k}) - 
B_s(2^
{k-1})$ then there is a $y \in X_0^{k}$ such that $d(x,y) \leq \ep 2^{k} < \ep 
2^
{k+1}$.  Otherwise $x \in B_s(2^{k-1})$.  By assumption, there is an $\hat{x} 
\in X_0^
{k-1}$
such that $d(x,\hat{x})\leq \ep2^{k}$, and by construction, there is a $y \in 
X_0^{k}$ such
that $d(y,\hat{x}) \leq \ep 2^{k}$.  Therefore $d(x,y) \leq \ep 2^{k} + 
\ep2^{k} 
= \ep2^
{k+1}$.
\end{proof}

\begin{lemma}\label{lem:embed_error}
 For all vertices $u$ and $v$, $\dist_G(u, v) \leq \dist_H(u, v) \leq 
\dist_G(u, 
v) + O(\ep)(\dist_G(s, u) + \dist_G(s, v))$
\end{lemma}

\begin{figure}[H]     
\centering  
\begin{subfigure}[t]{0.3\textwidth}
\centering
\includegraphics[width=0.65\textwidth]{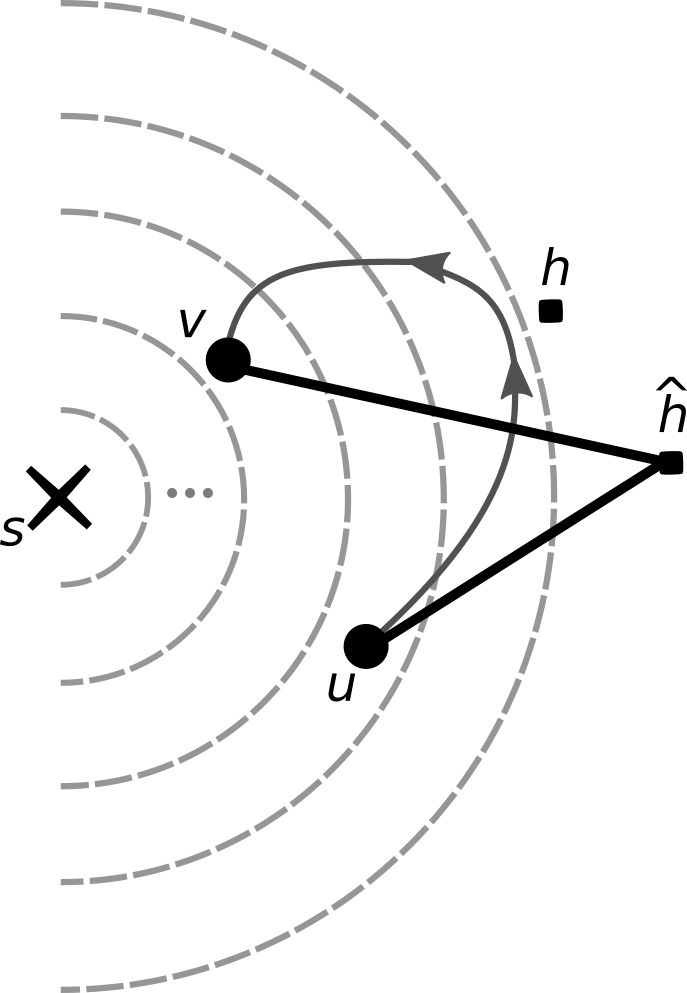}         
\caption{\centering $u$ and $v$ are both connected to $\hat h$}         
\label{fig:far}     
\end{subfigure}     
\qquad
\begin{subfigure}[t]{0.33\textwidth}
\centering
\includegraphics[width=0.50\textwidth]{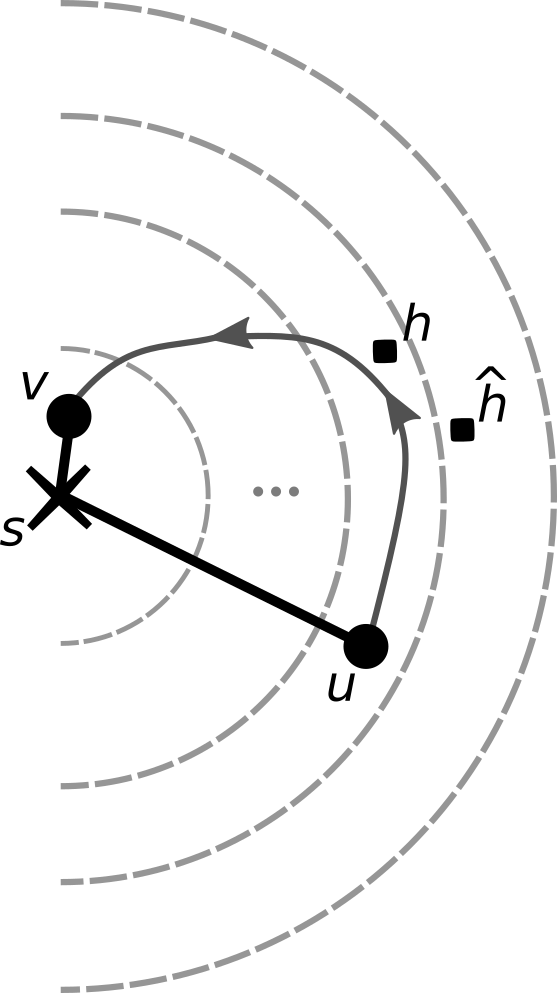}         
\caption{\centering $v$ is not connected to $\hat h$}         
\label{fig:close}     
\end{subfigure}
\caption{\label{fig:crossings}The shortest path between $u$ and $v$ in $G$ is 
indicated by the curved, directed lines. The path in the host graph is 
represented
by the
straight lines.}
\end{figure}

\begin{proof}
Consider two vertices $u$ and $v$. Let $T_u$ and $T_v$ denote the top-level 
towns that contain $u$ and $v$, respectively. 
There are two cases to consider. 

If $T_u = T_v$, Lemma~\ref{lem:towns} gives
$\dist_G(u, v) \leq diam(T_u) \leq \dist_G(T_u, V \setminus T_u)$, and 
therefore $diam(T_u) \leq \min\{\dist_G(s,u),\ \dist_G(s,v)\}$. 
Because $T_u = T_v$ is a top-level town, its embedding is given by 
Lemma~\ref{thm:embed}, which directly gives the desired bound. 

Otherwise $T_u \neq T_v$.  Without loss of generality, assume that $d_G(u, s) 
\geq d_G(v, s)$. We show that there exists some $X_0^k$ connected to $u$ with a 
vertex $\hat{h} \in X_0^k$ close to $P[u,v]$. 

By definition of the approximate core hubs, there exists $h \in 
X_{T_0}$ such that $d(h, P[u, v]) \leq \ep d(u, v)$. 
Moreover, $h \in B_s(2^{l(T_u) + 2})$:

\begin{table}[H]
\centering
\begin{tabular}{llr}
 $d(s, h)$ & $\leq d(s,u)+d(u,h)$ &\\
           & $\leq d(s, u) + (1+\ep)d(u,v)$&\\
           & $\leq d(s, u) + (1+\ep)\big(d(s,u) + d(s, v)\big)$ & by the 
triangle
inequality\\
	   & $\leq d(s, u) + (1+\ep) \cdot 2d(s,u)$ & since $d(u, s) \geq d(v, 
s)$\\
           & $\leq (3+2\ep)2^{l(T_u)}$&\\
           & $\leq 2^{l(T_u)+2}$&\\
\end{tabular}
\end{table}

Since $h \in X_{T_0} \cap B_s(2^{l(T_u)+2})$, then by 
Lemma~\ref{lem:hub_cover}, 
there is an $\hat{h}
\in 
X_0^{l(T_u)+2}$ such that $d(\hat{h}, h) \leq \ep 2^{l(T_u)+3}$. Since
$\log_2 \frac{1}{\ep} \geq 2$, $u$ is connected to $\hat{h}$ in the
host graph.

Depending on $v$, there remain two cases: either $v$ is connected to $\hat{h}$ 
(see Figure~\ref{fig:far}) or not (Figure~\ref{fig:close}). First, if $v$ is 
connected to $\hat h$ in the host graph, $d_H(v,\hat{h})
= d_G(v, \hat{h})$ (and the same holds for $u$). The triangle inequality gives 
therefore,
$$d_H(u, v) \leq d_G(u, \hat{h}) + d_G(v, \hat{h}) \leq \underbrace{d_G(u, h) + 
d_G(v, h)}_{\leq (1+2\ep)d_G(u, v)\ \text{\newline by definition of $h$}} 
+ \underbrace{2d_G(\hat{h}, h)}_{\leq 2\ep 2^{l(T_u)+3} = O(\ep)d(s, u)}$$

Since $d_G(u, v) \leq d_G(s, u) + d_G(s, v)$, we can conclude that, 
 $$d_H(u, v) \leq d_G(u,v) + O(\ep) (d_G(s, u) + d_G(s, v))$$ 
\medskip

Otherwise, $v$ is not connected to $\hat{h}$. That means that either 
$l(T_u)+2 < l(T_v)$ or $l(T_u)+2 > l(T_v) + \log_2 \frac{1}{\ep}$.  We exclude 
the
first case by noting that since the diameter of a town is less than its 
distance 
to
the depot, $d_G(v, s) \leq d_G(u, s)$ implies that $l(T_v) \leq l(T_u) + 1$.  
The
second case implies that $d_G(s, u) \geq O(\frac{1}{\ep}) d_G(s, v)$. 
Since the host graph connects the source $s$ to all the vertices, 
$d_H(u, v) \leq d_G(s, u) + d_G(s, v) \leq d_G(u, v) + 2d_G(s, v) \leq
d_G(u,v) + O(\ep) (d_G(s, u) + d_G(s, v))$.
\end{proof}

\bigskip

\bigskip
\subsection{Tree Decomposition}\label{ssec:treedec}
We present here the construction of a tree decomposition $D$ of the host graph 
with a bounded width.

For each $k>0$ let $\mathcal{B}_k = \bigcup\limits_{i=k-1}^{k+\log_2(1/ \ep)} 
X_0^i$. For a top-level town $T$, the tree decomposition $D$ connects the 
decomposition $D_T$ given by Lemma~\ref{thm:embed} to the bag 
$\mathcal{B}_{l(T)}$. Moreover, we add all vertices that appear in 
$\mathcal{B}_{l(T)}$ to all bags in the tree $D_T$. Finally, for every $k$ we 
connect $\mathcal{B}_k$ to both $\mathcal{B}_{k-1}$ and $\mathcal{B}_{k+1}$ in 
$D$. (See Figure~\ref{fig:town_hub_embed}.)

\begin{lemma}
 $D$ is a valid tree decomposition of the host graph $H$.
\end{lemma}

\begin{proof}
For $D$ to be a valid tree decomposition of $H$, it has to satisfy the three 
properties listed in Section~\ref{sec:prelims}.

First, because the top-level towns are a partition of the vertices, each 
vertex appears in some tree decomposition $D_T$. The union of all bags is 
therefore $V(H)$. 

Next, let $(u, v)$ be an edge of $H$.  There are two cases to consider: if $u$ 
and $v$ are in the same top-level town, Lemma~\ref{thm:embed} ensures that 
$u$ and $v$ appear together in some bag. Otherwise, as the top-level towns are 
disjoint, one of $u$ or $v$ is a hub connected to the other. 
Without loss of generality assume that $v$ is a hub of $X_0^k$ for some $k \in 
\{l(T_u), ..., l(T_u)+\log_2 \frac{1}{\ep}\}$. In this case, $v \in 
\mathcal{B}_{l(T_u)}$, so $v$ is added to all the bags of $D_{T_u}$, and in 
particular is in some bag that contains $u$. 

Finally, let $v$ be a vertex that appears in two different bags. If the 
two bags are in the tree decomposition of the same top-level town $T$, 
Lemma~\ref{thm:embed} ensures that the bags are connected in $D_T$ and thus 
also in $D$. Otherwise, as the top-level towns are disjoint, $v$ must be a hub. 
Consider all nets $X_0^k$ containing $v$.  Any bag $\mathcal{B}_\ell$ 
containing such a net also contains $v$. Let $\mathcal{I} = \{k | v \in 
X_0^k\}$. We prove that $\mathcal{I}$ is an interval, and therefore that the 
bags $\mathcal{B}_\ell$ are connected. Let $i = \min (\mathcal{I})$ and $j = 
\max (\mathcal{I})$. As $v \in X_0^i$, it must be that $v \in B_s(2^i)\subseteq 
B_s(2^{i+1}) \subseteq ... \subseteq B_s(2^j)$. Repeatedly applying the 
property $X_0^k \cap B_s(2^{k-1}) \subseteq X_0^{k-1}$ proves that for all $k 
\in \{i, i+1,... j\},\ v \in X_0^k$. Therefore $\mathcal{I}$ is an interval, 
and 
the bags $\mathcal{B}_\ell$ such that $v \in \mathcal{B}_\ell$ are connected. 
Finally, we show that interval $\mathcal{I}$ includes $\mathcal{B}_l(T_v)$.  
Since
$v$ is a hub, $v \in X_0^{l(\{v\})}$.  By Lemma~\ref{lem:towns}, $ d(v,s) > 
diam(T_v)$, so
$l(T_v) - 1\leq l(\{v\}) \leq l(T_v)$, and therefore $v\in \mathcal{B}_l(T_v)$. 
Since
the vertices of $\mathcal{B}_l(T_v)$ are added to every bag in  
$D_{T_v}$, the bags containing $v$ form a connected subtree of $D$.
\end{proof}

\begin{lemma}\label{lem:bag_size}
 For all $k$, $|X_0^k| \leq (\frac{2}{\ep})^\theta$.
\end{lemma}
\begin{proof}
 Since $X_0^k$ is a subset of $X_{T_0}$, it has doubling dimension 
$2\theta$ (see Lemma~\ref{lem:approx_core_hub}). Since $X_0^k$ is a $\ep 
2^k$-net, the smallest 
distance between two hubs in $X_0^k$ is at least $\ep 2^k$.  
 Moreover, since $X_0^k \subseteq B_s(2^k)$, the longest distance between two 
hubs is at most $2\cdot2^k$, therefore, $X_0^k$ has an aspect ratio of at most 
$\frac{2}{\ep}$. The bound used in Lemma~\ref{lem:tw} on the cardinality of a 
set using its aspect ratio and its doubling dimension concludes the proof.
\end{proof}

\begin{lemma}\label{lem:widthD}
 The tree decomposition $D$ has bounded width.
\end{lemma}
\begin{proof}
Bag $\mathcal{B}_i$ is the union of $\log_2 \frac{1}{\ep}+2$ sets $X_0^k$.
Lemma~\ref{lem:bag_size} gives $|X_0^k| \leq (\frac{2}{\ep})^\theta$, therefore 
$|\mathcal{B}_i| \leq (\log_2(\frac{1}{\ep})+2) (\frac{2}{\ep})^\theta$. 
Moreover,
by Lemma~\ref{lem:tw}, each bag of the $D_T$ decompositions has a cardinality 
bounded by $O((\frac{2}{\ep})^\theta \log_{\frac{c}{4}} \frac{1}{\ep})$. 
Therefore, since each bag of the decomposition $D$ is either a bag 
$\mathcal{B}_i$ for some $i$ or is formed by adding a single bag 
$\mathcal{B}_i$ 
to some bag of a $D_T$ decomposition, its size is bounded. Therefore $D$ has a 
bounded width. 
\end{proof}

\bigskip

\section{Capacitated Vehicle Routing}\label{sec:vehicle_routing}

\subsection{PTAS for Bounded Highway Dimension}
The {\sc Capacitated Vehicle Routing} problem for some graph $G$, demand 
function
$\rho:V\rightarrow[1,2,...,Q]$, \emph{depot} vertex $s \in
V$ and capacity $Q > 0$ seeks a set
of tours of minimal total length that collectively visit all clients (vertices 
with
positive client demand), such
that each tour contains $s$ and covers at most $Q$ units of client demand.  
In this section, we apply Theorem~\ref{thm:routing_embed} to 
{\sc Capacitated Vehicle Routing}, for graphs of bounded highway dimension 
$\hdimension$ and fixed capacity $Q$. 

The algorithm works as follows.  The input graph $G$ is embedded into a host 
graph
$H$ of bounded treewidth using the embedding given in 
Theorem~\ref{thm:routing_embed}.
The algorithm then optimally solves the {\sc Capacitated Vehicle Routing} 
problem
with capacity
$Q$ for $H$, using the dynamic programming algorithm given in 
Section~\ref{sec:dp}.
 The solution for $H$ is then \emph{lifted} to a solution in $G$: for each tour 
in
 the solution for $H$, a tour in $G$ that visits the same clients in the same 
order
 is
 added to the solution for $G$.

We show that the embedding given 
in Theorem~\ref{thm:routing_embed} is such that an optimal solution in the host 
graph $H$ gives a $(1+\ep)$ solution in $G$.  Furthermore, the embedding 
ensures that $H$ has small treewidth, allowing {\sc Capacitated Vehicle 
Routing} 
to be solved exactly in polynomial time using dynamic programing. Putting these 
together gives Theorem~\ref{thm:cvr_PTAS}, restated here for convenience.

\begin{restate}{\bf \ref{thm:cvr_PTAS}.}
For any $\epsilon>0$, $\hdimension > 0$ and $Q>0$,
  there is a polynomial-time algorithm that, given an instance
  of {\sc Bounded-Capacity Vehicle Routing} in which the capacity is $Q$
  and the graph has highway dimension $\hdimension$,
  finds a solution whose cost is at most $1+\epsilon$ times optimum.
\end{restate}

\bigskip
Given an embedding with the properties described in 
Theorem~\ref{thm:routing_embed}, all that remains in proving 
Theorem~\ref{thm:cvr_PTAS} is showing how to solve {\sc Capacitated Vehicle 
Routing} optimally on the host graph $H$ and proving that such an optimal 
solution has a corresponding \emph{near-optimal} solution in $G$. We do so in 
the following two lemmas.

\begin{lemma}
Given a graph with bounded treewidth $\omega$ and a capacity $Q>0$, {\sc 
Capacitated Vehicle Routing} can be solved optimally in $n^{O(\omega Q)}$ time.
\end{lemma}

\begin{proof}
See Section~\ref{app:cvr_dp}
\end{proof}

\begin{lemma}\label{lm:cvr_approx}
 For an embedding with the properties given by Theorem~\ref{thm:routing_embed}, 
the cost of an optimal solution in the host graph $H$ is within a 
$(1+O(\ep))$-factor of the cost of the optimal solution in the guest graph $G$.
\end{lemma}

\begin{proof}
 Let $\OPT_H$ be the optimal solution in the host graph $H$ and $\OPT_G$ be the 
optimal solution in $G$. A solution is described by the order in which 
the clients and the depot are visited: $(u, v) \in S$ indicates that the 
solution $S$ visits the client $v$ immediately after visiting $u$.
We want to prove that $\cost_G(\OPT_H) \leq (1+O(\ep))\cost_G(\OPT_G)$. 

 First, since $\dist_G \leq \dist_H$, $\cost_G \leq \cost_H$. 
Second, the solution $\OPT_G$ is also a solution in the host graph $H$, since 
the vertices of $G$ and $H$ are the same. So, by definition of $\OPT_H$, 
$\cost_H(\OPT_H) \leq \cost_H(\OPT_G)$. It is therefore sufficient to prove 
that $\cost_H(\OPT_G) \leq (1+ O(\ep))\cost_G(\OPT_G)$.

 By definition of cost, $\cost_H(\OPT_G) = \sum\limits_{(u, v)\in \OPT_G} 
\dist_H(u,v)$. Applying Theorem~\ref{thm:routing_embed} gives 
 $$\cost_H(\OPT_G) \leq  \sum\limits_{(u, v)\in \OPT_G} \dist_G(u, v) 
+ O(\ep)(\dist_G(s, u) + \dist_G(s, v))$$
 The right side of the inequality can be rewritten as  
$$\underbrace{
\sum\limits_{(u,v)\in \OPT_G} \dist_G(u,v)}_{=\ \cost_G(\OPT_G)} \ \ 
+\ \ \underbrace{
O(\ep) \sum\limits_{(u, v)\in \OPT_G}\dist_G(s, u)+ \dist_G(s, v)}
_{=\ O(\ep)\sum\limits_{v \in Z} 2\dist_G(s, v)\newline\ \leq\ 
O(\ep)Q\cost_G(\OPT_G) \ \ (*)}$$

To get the inequalities $(*)$, it is enough to remark that $\OPT_G$ visits 
every 
client exactly once and then to apply Lemma~\ref{lem:lw_bound}. As $Q$ is
constant, the whole inequality becomes $$\cost_H(\OPT_G) \leq \cost_G(\OPT_G) + 
O(\ep)\cost_G(\OPT_G) = (1+O(\ep))\cost_G(\OPT_G)$$
\end{proof}

\subsection{Generalization to Routing with Penalties}\label{sec:generalization}
\label{sec:penalties}
The {\sc Capacitated Vehicle Routing with 
Penalties} is a natural generalization of {\sc Capacitated Vehicle
Routing} in which a penalty is specified for each client, and the
solution can omit some clients (and pay their penalties).
The embedding proposed previously can be used to solve it. First, the 
dynamic program for graphs of bounded treewidth can be adapted to solve 
this problem optimally in such graphs. The only change to make
is that instead of visiting a client, the algorithm can chose to pay the 
penalty. It remains to prove that an optimal solution in the host graph is 
close to an optimal solution in the guest graph.

\begin{lemma}
  The optimal solution to {\sc Capacitated Vehicle Routing with Penalties} in 
the host graph has a cost at most $(1+\ep)\cost(\OPT_G)$
\end{lemma}
\begin{proof}
The clients can be divided into two sets $U$ and $W$: the optimal solution in 
$G$ visits every vertex in $U$ and pays the penalty for the ones in $W$. 
Applying Lemma~\ref{lem:lw_bound} to the set $U$, gives the following: 
$$\cost(\OPT_G) \geq \frac{2}{Q}\sum\limits_{v \in U}\dist(v, s) + 
\sum\limits_{v \in W} p(v)$$

With this lower bound, the proof of Lemma~\ref{lm:cvr_approx} can be adapted to 
handle penalties, giving $\cost_H(\OPT_G) \leq (1+O(\ep))\cost_G(\OPT_G)$. 
The conclusion is similar to the one of Lemma~\ref{lm:cvr_approx}: 
$\cost_G(OPT_H) \leq (1+O(\ep))\cost_G(\OPT_G)$.
\end{proof}

\section{Dynamic Program for Capacitated Vehicle 
Routing}\label{sec:dp}\label{app:cvr_dp}

In this section, we present a dynamic program running in $n^{O(\omega Q)}$ to 
solve {\sc Capacitated Vehicle Routing} for capacity $Q$ on graphs with 
treewidth $\omega$. Given a tree decomposition, $D$, choose an arbitrary bag
to be the root, and for each bag $b$ of the decomposition let \emph{cluster} 
$C_b$ be the union of the bags descending from $b$ in the tree decomposition, 
minus the elements of $b$ itself. The bag $b$ forms a boundary between cluster 
$C_b$ and $V \setminus C_b$. 

A configuration in the dynamic program describes how a solution interacts with 
a cluster: for each vertex $v$ in the boundary $b$ of the cluster, and for each 
possible capacity $q \leq Q$, the configuration specifies $I_{v, q}$ and $O_{v, 
q}$ which are respectively the number of tours that enter and exit $C_b$ by 
vertex $v$ and that have visited exactly $q$ clients at the moment they reach 
$v$. We refer to this as the \emph{flow} in and out of $C_b$ at $v$. These 
values are sufficient to recover the intersection of the solution with the 
cluster: connecting each entering tour with an exiting one, at minimal cost, 
gives the optimal solution.

To simplify the dynamic program, we first convert $D$ into a 
\emph{nice} tree decomposition with $O(\omega n)$ bags. This can be done in 
polynomial time, while preserving the width~\cite{param_algo}. In a nice 
tree decomposition, each leaf bag contains a single vertex and each non-leaf bag
of the decomposition is one of
three types: 
\begin{itemize}[noitemsep]
 \item An \emph{introduce} bag $b$ has one child $b'$, such that $b = b'
\cup \{v\}$ for some vertex $v \notin b'$.  The vertex $v$ is introduced at $b$.
 \item A \emph{forget} bag $b$ has one child $b'$, such that $b = b'
\setminus \{v\}$ for some vertex $v \in b'$. The vertex $v$ is forgotten at $b$.
 \item A \emph{join} bag $b$ has two children $b_1$ and $b_2$ such that $b = b_1
 = b_2$.
\end{itemize}

Moreover, as observed in \cite{param_algo}, the third property of a tree 
decomposition ensures that each vertex can be forgotten only once. 

Furthermore, we assume the forget bag for the depot occurs at the root of the
decomposition.  If not, $s$ can be added to every bag in the tree, and the 
leaves
extended.  This results in a nice tree decomposition with at most twice as many 
bags
while adding at most one to the width.

A tour can be uncrossed to avoid crossing the same vertex in the same 
direction twice. As there are at 
most $n$ different tours, $I_{v,q} \leq n$ and $O_{v, q} \leq n$, so there are 
$n
^{O(\omega Q)}$ possible configurations per bag. Since there
are $O(\omega n)$ bags in the nice tree decomposition, there are a total of $n 
^{O
(\omega Q)}$ different configurations.

The algorithm runs bottom-up: given a configuration for each child node, it 
finds all possible \emph{compatible} configurations for the parent node. Each 
different type of bag of the nice tree decomposition requires a particular 
treatment. 

\medskip

For a leaf bag, $b$, containing vertex $u$, $C_b$ is empty, so trivially there 
are no
tours entering or exiting $C_b$. For configurations with $I_{u,q} = O_{u, q} = 
0$
for all $q$, the algorithm stores the cost zero.

\medskip

For a forget bag, the parent bag $b$ is equal to its child bag $b'$ minus some 
vertex $u$. For each child bag configuration, the algorithm considers all ways 
to form a compatible parent bag configuration by rerouting $u$'s flow and, if 
$u$ is a client, covering its demand, $\rho(u)$.  For each resulting parent bag 
configuration, the dynamic program stores the cost only if it is less than the 
current value stored for that configuration. After considering all child bag 
configurations and ways of forming a parent bag configuration, the values 
stored in the table are guaranteed to be optimal.  Consider some configuration
for the child bag. First, if $u$ is a client, one tour is selected to visit it. 
There are three cases.  If the tour crosses into $C_{b'}$ after visiting
$u$, the algorithm chooses a capacity $q \geq \rho(u)$ and
makes the following changes to the flow at $u$: 
$$I_{u, q} \rightarrow I_{u, q} - 1,\ \ I_{u, q-\rho(u)} \rightarrow 
I_{u, q-\rho(u)}+1$$
There are at most $Q$ such choices.  

If the tour crosses out of $C_{b'}$ after visiting
$u$, the algorithm chooses a capacity $q \leq Q-\rho(u)$ and
makes the following changes to the flow at $u$: 
$$O_{u, q} \rightarrow O_{u, q} - 1,\ \ O_{u, q+\rho(u)} \rightarrow 
O_{u, q+\rho(u)}+1$$
There are at most $Q$ such choices.  

Otherwise, the tour segment that visits $u$ does 
not cross into $C_{b'}$.  The algorithm chooses $v_1,v_2 \in b$ and $q \leq 
Q-\rho
(u)$, makes the following changes to the flow at $v_1$ and $v_2$:
$$I_{v_1, q} \rightarrow I_{v_1, q} + 1,\ \ O_{v_2, q+\rho(u)} \rightarrow 
O_{v_2,
q+\rho(u)}+1,$$
and adds $d(v_1,u)+d(u,v_2)$ to the intermediate configuration cost.  There are 
$\omega^2Q$ such choices. The algorithm then reroutes all flow through $u$ to 
some vertex in the parent bag, $b$. The algorithm chooses, for each vertex $v$ 
of $b$ and each capacity, the number of the tours that enter (resp. exit) 
$C_{b'}$ though $u$ directly from (resp. to) $v$. Each such tour adds a cost of 
$\dist(u, v)$ to the intermediate configuration cost. There are $O(n^{2\omega 
Q})$ 
such choices.  Thus, for each child configuration there are 
$O(\omega^2Qn^{2\omega Q})$
choices, giving an $n^{O(\omega Q)}$ overall runtime for each forget bag.

\medskip
For an introduce bag, the parent bag is equal to its child bag plus some vertex
$u$. Since the child bag forms a boundary between the inside and outside of the 
cluster, no tour can cross directly into the cluster via $u$, as it must first 
cross some vertex of the child bag.  Therefore the only compatible parent 
configurations are those that have no tours crossing at $u$. So for every 
parent 
configuration, if $I_{u,q} = O_{u, q} = 0$ for all $q$, the algorithm stores 
the 
cost of the corresponding child configuration, namely the configuration that 
results by removing $u$. Otherwise the cost is $\infty$.

\medskip
For a join bag, the parent bag has two child bags identical to itself. 
Lemma~\ref{lem:oracle} presents an oracle that tells, in constant time, the 
minimal cost needed to form parent configuration $(I^0, O^0)$ given child 
configurations $(I^1, O^1)$ and $(I^2, O^2)$, with an infinite cost if the 
configurations are not compatible. The algorithm tries all combinations of 
configurations: the complexity of this step is $n^{O(\omega Q)}$.

\medskip
Since each vertex will appear exactly once in a forget bag, each client will be 
visited exactly once. The overall complexity is $n^{O(\omega Q)}$, as claimed. 
The algorithm considers all possible solutions and outputs the minimal one, so 
the resulting cost is optimal.

\begin{lemma}\label{lem:oracle}
 For each join bag $b$, it is possible to compute, in $O(n^{6 \omega Q})$ time, 
a table $\mathcal{T}_b$ such that $\mathcal{T}_b[(I^0, O^0), (I^1, O^1), (I^2, 
O^2)]$ is the minimal cost to connect child configurations $(I^1, O^1)$ and 
$(I^2, O^2)$ to form parent configuration $(I^0, O^0)$ of $b$.
\end{lemma}

\begin{proof}
 We design a dynamic program to compute this table.  The base cases are when 
$I^0 = I^1 + I^2$. If $O^0 = O^1 + O^2$ the cost is 0, since the configurations 
are therefore compatible. Otherwise the cost is $\infty$, because it is not 
possible to balance incoming and outgoing flow.
 
For the recursion step, assume $I^0 \ne I^1 + I^2$. Pick the first pair 
$(u, q)$ such that $I^1_{u, q} + I^2_{u, q} - I^0_{u, q} = x \ne 0$. If $x < 
0$, the incoming flow at $u$ with capacity $q$ is bigger in $b$ than in its 
child bags. Since this is not possible, the cost is $\infty$. Otherwise, some 
flow entering Cluster 1 comes from Cluster 2 (or vice versa). Suppose this flow 
exits Cluster 2 at vertex $v$: it means that $$\mathcal{T}_b[(I^0, O^0), (I^1, 
O^1), (I^2, O^2)] = \mathcal{T}_b[(I^0, O^0), (\hat I^1, O^1), (I^2, \hat O^2)] 
+ \dist(u, v)$$ where $\hat I^1 = \set{I^1, I^1_{u, q}-1}$ and $\hat O^2 = 
\set{O^2, O^2_{v, q}-1}$. By this equation, the algorithm connects one segment 
exiting Cluster 2 at $v$ with capacity $q$ to a segment entering Cluster 1 at 
$u$. The value of $\mathcal{T}_b[(I^0, O^0), (I^1, O^1), (I^2, O^2)]$ can 
therefore be computed in $\omega$ steps, by applying the above equality for 
each 
vertex $v$ of the boundary and storing the minimum value. This computation 
requires $O(\omega Q)$ operations to find the pair $(u, q)$, and then 
$O(\omega)$ operations to compute the value of the table. The recursion step 
therefore requires $O(\omega Q)$ time.

As there are $O(n^{6 \omega Q})$ states for this DP, the overall complexity is 
therefore $O(\omega Q n^{6 \omega Q}) = O(n^{6 \omega Q})$, concluding the 
proof.
\end{proof}

\section{Embedding for Multiple Depots}\label{sec:multiple}

We present in this section how to extend Theorem~\ref{thm:routing_embed} 
and apply it to several problems.

\subsection{Theorem}

\begin{restate}{\bf \ref{thm:multi}} There is a function
  $f(\cdot,\cdot,\cdot)$ such that,
 for every $\ep > 0$, graph $G$ of highway dimension $\hdimension$
and set $S$ of vertices of $G$, there exists a graph $H$ and an embedding 
$\phi(\cdot)$ of $G$ into $H$ such that 
\begin{itemize}[noitemsep]
 \item $H$ has treewidth $f(\hdimension, |S|, \ep)$, and
 \item for all vertices $u$ and $v$,
$$\dist_G(u, v) \leq \dist_H(\phi(u), \phi(v)) \leq 
(1+O(\ep))\dist_G(u, v) + \ep\min(\dist_G(S, u), \dist_G(S, v))$$
\end{itemize}
\end{restate}

We slightly modify the embedding of Theorem~\ref{thm:routing_embed} in that 
purpose.  We assume that the vertices of $S$ do not appear in non-trivial 
towns. 
This assumption is \emph{safe} because, using $S$, we can apply the 
modification 
of
Lemma~\ref{lem:depot_safety} to satisfy this assumption without asymptotically 
changing
the diameter or size of the graph. Note that the modification preserves all 
distances
from the original input graph but increases the highway dimension to 
$\hdimension
+ |S|$.

The algorithm computes the town decomposition with 
respect to the shortest-path covers, and embeds the top-level towns using 
Lemma~\ref{thm:embed}. 
By analogy with Section~\ref{sec:main_embed}, we define the set $X_0^i$ to be a 
$\ep2^i$-net of $\cup_{s \in S} B_{s}(2^i)$ (and we ensure moreover that the 
$X_0^i$ are nested). We also modify the definition of $l(\mathcal{X})$: for a 
set $\mathcal{X}$, $l(\mathcal{X}) =
\lceil \log_2 ( \max_{v \in \mathcal{X}} \dist(S, v) ) \rceil$. Following 
Section~\ref{sec:main_embed}, the host graph connects every vertex $v$ of a 
town 
$T$ to every hub $h$ in $X_0^{l(T)}, \ldots, X_0^{l(T) + \log_2(1/ \ep)}$ with 
an edge of length $d_G(v,h)$.

We now prove that this embedding has the properties of 
Theorem~\ref{thm:multi}. We use $H$ to denote the host graph produced by the 
embedding. First, we prove the first point: the treewidth is bounded.

\bigskip
\begin{proof}
 Let $\theta_S$ be the doubling dimension of the union of the approximate core 
hubs
 with $S$. The shortest-path covers 
are locally $(\hdimension+|S|) \log (\hdimension+|S|)$-sparse \cite{abraham11}, 
therefore
Lemma~\ref{lem:approx_core_hub} gives, 
$\theta_S=O\Big(\log\big((\hdimension+|S|)^2 \log (\hdimension+|S|)\log(1 / 
\ep) 
+
|S|\big)\Big)$, where the final $|S|$ term comes from the extra balls required 
to
cover $S$. The proof of
Lemma~\ref{lem:bag_size} directly gives that $|X_0^i| \leq 
|S|(\frac{2}{\ep})^{\theta_S}$. Finally, following the proof of 
Lemma~\ref{lem:widthD}, the host graph has a treewidth bounded by a function of 
$\hdimension,\ |S|$ and $\ep$.
\end{proof}

\bigskip
We now prove the distortion bound.
\begin{proof}
 Let $u$ and $v$ be two points of the metric space and $h$ be the approximate 
core hub such that $\dist_G(u, h) + \dist_G(v,h) \leq (1+O(\ep))\dist_G(u, v)$.
Let $s_u,\ s_v$ and $s_h$ denote the points of $S$ closest to $u, v$ and $h$. 
The proof is divided into three parts, according to the distances between 
$l(h)$, $l(T_u)$ and $l(T_v)$.

\begin{figure}[!ht]
  \centering
  \begin{subfigure}[t]{0.25\textwidth}
    \includegraphics[width=\textwidth]{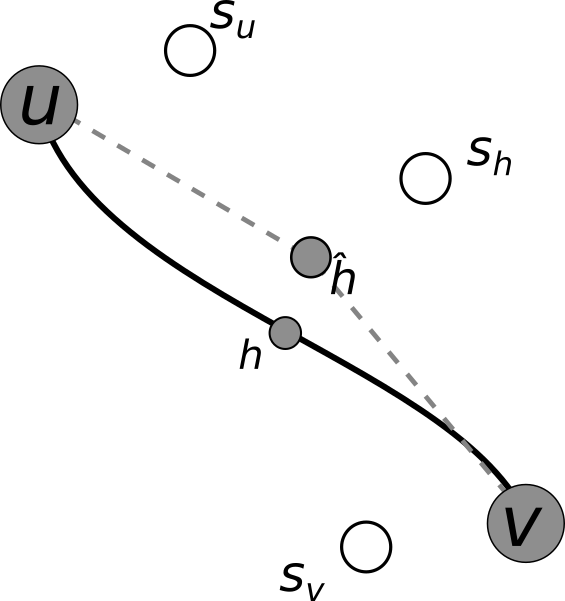}
    \caption{$u$ and $v$ are adjacent to $\hat h$}
    \label{fig:multi1}
  \end{subfigure}
  \quad\vline\quad
  \begin{subfigure}[t]{0.25\textwidth}
    \includegraphics[width=\textwidth]{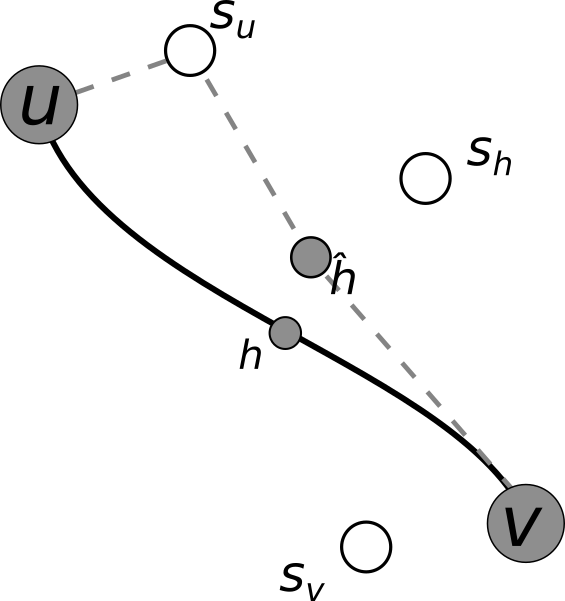}
    \caption{only $v$ is adjacent to $\hat h$}
    \label{fig:multi2}
  \end{subfigure}
  \quad\vline\quad
  \begin{subfigure}[t]{0.25\textwidth}
    \includegraphics[width=\textwidth]{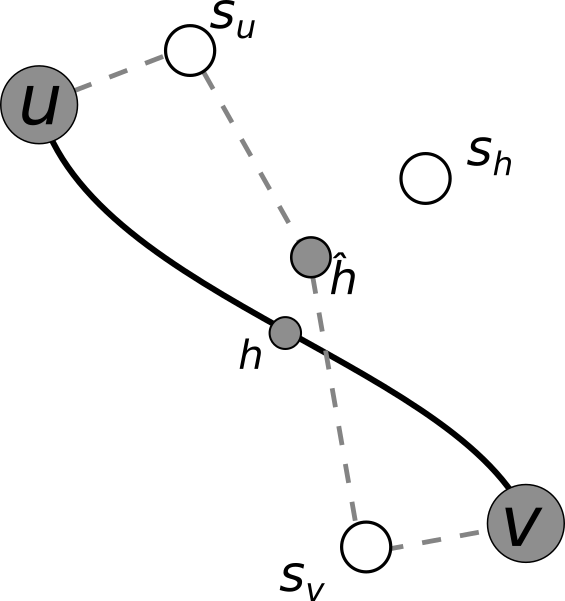}
    \caption{$u$ and $v$ are not adjacent to $\hat h$}
    \label{fig:multi3}
  \end{subfigure}
    \caption{  \label{fig:multi}The shortest path between $u$ and $v$ is  
approximated in the host graph by the dashed line. In case (a), the distance 
from $u$ and $v$ to their centers is large compared to the distance between $h$ 
and $s_h$; in case (b), only the distance from $v$ to $s_v$ is large, and in 
case (c) the distance from $h$ to $s_h$ is larger than the other two.}
\end{figure}

We first prove an inequality that is used several times: 
\begin{equation}
\label{eq:hbd} d_G(h, s_h) \leq (1+O(\ep))d_G(u,v)+\min( d_G(u, s_u), d_G(v, 
s_v))
\end{equation}

The definition of $s_h$ leads indeed to $d_G(h, s_h) \leq d_G(h, s_u)$ and 
using the triangle inequality we obtain $d_G(h, s_h) \leq d_G(h, u)+d_G(u, 
s_u)$. By definition, $h$ is near the shortest path between $u$ and $v$: this 
gives the desired bound for $u$ (the same holds for $v$).

\bigskip
Consider three cases, illustrated in Figure~\ref{fig:multi}. Suppose that 
both $l(h) \leq l(T_u) + \log_2(1/ \ep)$ and $l(h) \leq 
l(T_v) + \log_2(1/ \ep)$ (see Figure~\ref{fig:multi1}). Let $\hat h$ be the 
point in $X_0^{l(h)}$ closest to $h$: by definition of a net, $d_G(h, \hat 
h) \leq \ep2^{l(h)} \leq 2\ep d_G(h, s_h)$; by construction of 
the embedding, $\hat h$ is adjacent to $u$ and $v$. In this 
case, we have 
$$d_H(u,v) \leq d_H(u, \hat h)+ d_H(\hat h, v) \leq d_G(u, \hat h)+ d_G(\hat h, 
v) \leq d_G(u, h) + d_G(v, h) +2d_G(h, \hat h)$$
We infer from the definition of $h$ and $\hat h$ that $d_H(u,v) \leq 
(1+O(\ep))d_G(u, v) + 4\ep d_H(h, s_h)$ and using Equation~\ref{eq:hbd} 
$$d_H(u,v) \leq (1+O(\ep))d_G(u,v) + O(\ep)\min(d_G(u, S), d_G(v, S))$$.

Then suppose that $l(h) > l(T_u) + \log_2(1/ \ep)$ but $l(h) \leq l(T_v) + 
\log_2(1/ \ep)$ (see Figure~\ref{fig:multi2}). It means that $u$ is not 
adjacent to $\hat h$ but $v$ is. It means in particular that $\dist_G(s_h, h) 
> \frac{1}{\ep}d_G(s_u, u)$. The shortest-path between $u$ and $v$ is therefore 
approximated in the host graph by the path $u$, $s_u$, $\hat h$, $v$. 
The edges along this path have the length as in $G$, therefore 
$d_H(u, v) \leq d_H(u, s_u) + d_H(s_u, \hat h) 
+ d_H(\hat h, v) \leq d_G(u, s_u) + d_G(s_u, \hat h) + d_G(\hat h, v)$.
We now apply the triangle inequality in $G$: $d_G(s_u, \hat h) \leq d_G(s_u, 
u) + d_G(u, \hat h)$. Using the former inequality and previously-derived bounds 
gives
$$d_H(u, v) \leq 2 d_G(u, s_u) + d_G(u, \hat h) + d_G(\hat h, v) \leq 2\ep 
d_G(h, s_h) + d_G(u, h) + d_G(h, v) + 2d_G(\hat h, h)$$
Recall that $d_G(h, \hat h) \leq 2\ep d_G(h, s_h)$ and $d_G(u, h) + d_G(h, v) 
\leq (1+\ep)d_G(u,v)$. Using this and Equation~\ref{eq:hbd} finally gives
$$d_H(u, v) \leq (1+\ep)d_G(u, v) + 4\ep d_H(h, s_h) \leq (1+O(\ep))d_G(u,v) + 
O(\ep)\min(d_G(u, S), d_G(v, S))$$.

Finally, suppose that $l(h) > l(T_u) + \log_2(1/ \ep)$ and $l(h) > l(T_v) + 
\log_2(1/ \ep)$ (see Figure~\ref{fig:multi3}). It means in particular that 
neither $u$ nor $v$ is adjacent to $\hat h$. In this case, the shortest 
path between $u$ and $v$ is approximated in the host graph by the path $u, s_u, 
\hat h, s_v, v$: using the same arguments as in the former case, we derive that 
$d_H(u, v) \leq (1+O(\ep)) d_G(u, v) + O(\ep) \min(d_G(u, S), d_G(v, S))$.
\end{proof}

\subsection{Applications} \label{sec:app}
\subsubsection*{{\sc Multiple-Depot Capacitated
Vehicle Routing}}
The first application we consider is for {\sc Multiple-Depot Capacitated
Vehicle Routing} with a constant number of depots. Let
$S$ denote the set of depots, and recall that $Z$ is the set of clients.  We 
assume
that any vertices added in the modification in Lemma~\ref{lem:depot_safety} do 
not
have any client demand.

Generalizing the algorithm from Section~\ref{sec:vehicle_routing} relies on  
generalizing the lower bound given in Lemma~\ref{lem:lw_bound} to $\frac{1}{Q} 
\sum \set{d(c, S)\ :\ c\in Z}$, as proved
in \cite{esa}. This lower bound
allows 
for an error of $\ep d(c, S)$ for each client $c$: the embedding of 
Theorem~\ref{thm:multi} can therefore be applied.

\begin{theorem}\label{thm:multiple-depot-vehicle-routing}
For any $\epsilon>0$, $\hdimension > 0$, $k$ and any $Q>0$,
  there is a polynomial-time algorithm that, given an instance
  of {\sc Multiple-Depots Capacitated Vehicle Routing} in which the capacity is 
$Q$, the number of depots is $k$ and the graph has highway dimension at most 
$\hdimension$, finds a solution whose cost is at most $1+\epsilon$ times 
optimum.
\end{theorem}

The proof that an optimal solution in the host graph gives an 
approximate solution on the original graph follows directly from 
Lemma~\ref{lm:cvr_approx}, and the DP presented in Section~\ref{sec:dp} can be 
extended easily: for a constant number of depots and a constant highway 
dimension, the embedding gives a constant treewidth.

\bigskip
\subsubsection*{ {\sc $k$-center}}
Another application is to get a fixed-parameter approximation (FPA) for {\sc 
$k$-center} in a graph $G$ with highway dimension $\hdimension$, i.e. an 
algorithm with running time $f(\hdimension, k)n^{O(1)}$.

The algorithm proceeds in two steps: first, computes a constant-factor 
approximation $S$ (see \cite{hochbaum} or \cite{gonzalez} for a 
2-approximation). Applying Theorem~\ref{thm:multi} to $G$ with the set $S$ 
gives 
a host graph. Finally, the algorithm runs a DP that gives a 
$(1+\ep)$-approximation of the optimal solution in the host graph (where any 
vertices
added in the modification in Lemma~\ref{lem:depot_safety} are not required to 
be 
covered).
We prove
that 
this solution is also a $(1+\ep)$-approximation of the optimal solution in the 
original graph.

\begin{lemma}\label{lem:kcenter-approx}
 A $(1+\ep)$-approximation of $k$-center in the host graph given by 
Theorem~\ref{thm:multi} is a  $(1+O(\ep))$-approximation of $k$-center in the 
original graph.
\end{lemma}

\begin{proof}
 Let $\OPT_H$ denote the optimal solution in the graph $H$. For each vertex 
$u$, let $c_u$ denote the closest center to $u$ in $\OPT_G$. We have the 
following: 
$$\cost_H(\OPT_G) = \max\limits_{u \in V(G)} d_H(u, c_u) \leq \max\limits_{u 
\in 
V(G)} (1+O(\ep))d_G(u, c_u) + O(\ep)\min(d_G(u, S), d_G(c_u, S))$$ 
This inequality can be rewritten 
$$\cost_H(\OPT_G) \leq (1+O(\ep))\max\limits_{u \in V(G)} d_G(u, c_u)\ \ \ +\ \ 
\ O(\ep)\max\limits_{u \in V(G)}d_G(u, S)$$

Since the set $S$ is a $O(1)$-approximate solution in $G$,
$\cost_H(\OPT_G) \leq (1+O(\ep)) \cost_G(\OPT_G) + O(\ep)\cost_G(\OPT_G) = 
(1+O(\ep))\cost_G(\OPT_G)$. By definition of $\OPT_H$, $\cost_H(\OPT_H) 
\leq \cost_H(\OPT_G)$ and therefore $\cost_H(\OPT_H) \leq 
(1+O(\ep))\cost_G(\OPT_G)$. That is, since the optimal solution in $H$ is an 
approximate solution in $G$, an approximate solution in $H$ is 
also an approximate solution in $G$.
\end{proof}

The complexity of finding a constant-factor approximation and of 
constructing the embedding is a polynomial in $n$ with fixed degree. The 
complexity of the DP given by Schild, Fox-Epstein and Klein~\cite{EpsteinKS} 
for 
a treewidth $tw$ is $O(n (\log n)^{tw})$ which is $O(n^{O(1)} tw^{2tw})$ 
following Lemma 1 in Katsikarelis et al.~\cite{katsikarelis}. As the treewidth 
only depends on the highway dimension $\hdimension$, $k$ and $\epsilon$, the 
FPA 
claims follows. 

\bigskip
\subsubsection*{{\sc $k$-median}}
The last application presented here is to get a FPA {\sc $k$-median}. The 
outline is the same as for {\sc $k$-center}: first compute a constant-factor 
approximation $S$ (see \cite{shmoys1997approximation}), then apply 
Theorem~\ref{thm:multi} using the set $S$ and finally compute an approximate 
solution in the host graph. The dynamic program for $k$-center can be adapted 
to 
solve $k$-median with the same complexity (again, any vertices
added in the modification in Lemma~\ref{lem:depot_safety} do not contribute to 
the
cost). The following lemma is
straightforward:

\begin{lemma}
 A $(1+\ep)$-approximation of $k$-median in the host graph given by 
Theorem~\ref{thm:multi} is a  $(1+O(\ep))$-approximation of $k$-median in the 
original graph.
\end{lemma}
The proof is indeed the same as for Lemma~\ref{lem:kcenter-approx}, replacing 
the max by a sum.

\begin{restate}{\bf \ref{thm:kctr}}  There is a function $f(\cdot, \cdot,
  \cdot)$ and a constant $c$ such that,
  for each of the problems {\sc $k$-Center} and {\sc $k$-Median},
  for any $\hdimension > 0, k > 0$
  and $\ep > 0$, there is an $f(\hdimension, k, \ep)n^{c}$ algorithm
  that, given an instance in which the graph has highway dimension at
  most $\hdimension$, finds a solution whose cost is at most $1+\ep$
  times optimum.
\end{restate}

\section*{Acknowledgements}
Thanks to Andreas Feldmann and Vincent Cohen-Addad for helpful discussions and 
comments.
\bibliographystyle{abbrv}
\bibliography{bib}

\appendix
\section{Definitions of highway dimension} \label{sec:highway-dimension}

The definition of highway dimension we use is the one given by Feldmann et 
al.~\cite{feldmann}. However, alternate definitions exist. We summarize 
here the differences between them that are discussed in Feldmann et al. The 
original definition comes from Abraham et al.~\cite{abraham10}, in 2010. Their 
work uses $c=4$, but interestingly they remark that ``one  could  use  constants
bigger than  4''. Nonetheless, Feldmann et al.~\cite{feldmann} shows that 
changing the constant is not innocuous: for any constant $c$, there is a graph 
with $n$ vertices that has highway dimension 1 with respect to $c$ and highway 
dimension $\Omega(n)$ with respect to any $c' > c$.

Another definition of highway dimension comes from a 2011 paper of Abraham et 
al.~\cite{abraham11}. Their definition differs from Definition~\ref{def:hg_dim}
in that they use $c=2$ and all shortest paths of length in $(r, 2r]$ that
\emph{intersect} the ball $B_{v}(2r)$ (not just the ones that stay inside the 
ball). This is a generalization of Definition~\ref{def:hg_dim} for $c=4$: 
a path of length at most $2r$ that intersects the ball $B_{v}(2r)$ is also
entirely contained in the ball $B_{v}(4r)$. As is, the results of Feldmann et 
al.
and, consequently, the ones presented in this paper cannot be generalized 
to this definition.

The last noteworthy definition of highway dimension was also 
introduced 
by
Abraham et al.~\cite{abraham} in a journal paper in 2016 (we use $h$ to denote 
this parameter, different from $\hdimension$). This definition is 
stricter
than the one of 2010 (and therefore the one of our paper), Feldmann et al.
show that if a metric has a highway dimension $h$ according to the 2016 
definition, it has a highway dimension $O(h^2)$ according to the 2010 
definition.

\section{Proof Sketches of Section~\ref{sec:prelims} 
Lemmas}\label{sec:town_proofs}

We now give proof sketches for Lemma~\ref{lem:towns} and 
Lemma~\ref{lem:town_decomp},
restated here for convenience.  See \cite{feldmann}
for full proofs.

\begin{figure}[!ht]
  \centering
  \begin{subfigure}[t]{0.25\textwidth}
    \includegraphics[width=\textwidth]{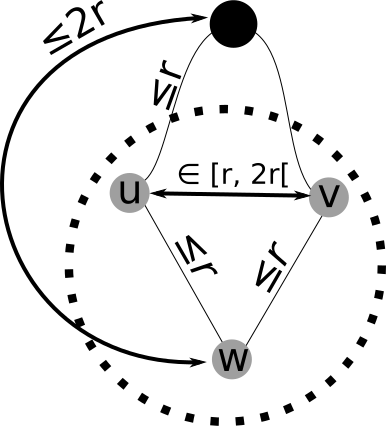}
    \caption{First contradiction: $d(u, v) \leq r$}
    \label{fig:towns-diam}
  \end{subfigure}
  \quad\vline\quad
  \begin{subfigure}[t]{0.25\textwidth}
    \includegraphics[width=\textwidth]{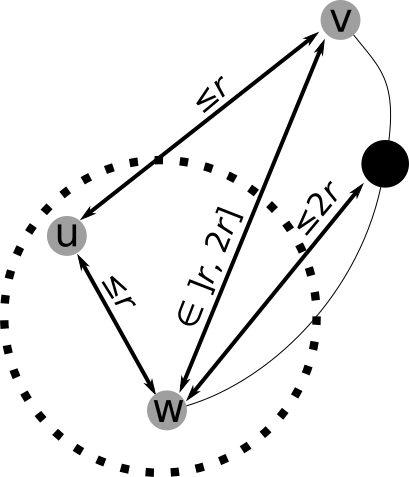}
    \caption{Second contradiction : $d(T, V \setminus T) > r$}
    \label{fig:towns-dist-ext}
  \end{subfigure}
    \caption{The shortest path cover is represented with black dots, the other 
points are grey dots. The dashed circles are towns.}
\end{figure}

\begin{restatelemma}{\bf \ref{lem:towns}.}[Lemma 3.2 in \cite{feldmann}]

If $T$ is a town at scale $r$, then
\begin{enumerate}[noitemsep]
\item $diam(T) \leq r$ and 
\item $d(T,V\setminus T) > r$ 
\end{enumerate}
\end{restatelemma}
\begin{proof}
 We give a sketch of the proof from \cite{feldmann}. For the first point 
(illustrated on Figure~\ref{fig:towns-diam}), let 
$u$ and $v$ be two vertices of the same town. It means that there exists $w$ 
such that $d(u, w) \leq r$ and $d(v, w) \leq r$ (by definition of a town). So 
$d(u, v) \leq 2r$. Suppose by contradiction that $d(u, v) > r$ : it means that 
$d(u, v) \in (r, 2r] \subseteq (r, cr/2]$ (because $c > 4$). Therefore there 
exists a point of the shortest-path cover for scale $r$ on the path between $u$ 
and
$v$: this point is at a distance at most $r$ from $u$ or $v$ (because $d(u, v) 
\leq 2r$), and therefore $2r$ from $w$, which contradicts the 
construction of the town. Therefore $d(u, v) \leq r$, and $diam(T) \leq r$.

We proceed similarly for the second point, illustrated in 
Figure~\ref{fig:towns-dist-ext}. Let $u \in T, v \in V \setminus T$, 
and suppose by contradiction that $d(u, v) \leq r$. The definition of $T$ gives 
a point $w$ such that $d(u, w) \leq r$, $d(v, w) > r$ and $d(w, SPC(r)) > 2r$. 
Combining the inequalities gives that $r < d(v, w) \leq d(u, v) + d(u, w) < 2r 
< cr/2$. By definition of the shortest-path cover, there is a hub $h$ on the 
shortest
path between $v$ and $w$, so $d(w, h) \leq d(w, v) \leq 2r$, which contradicts 
the choice of $w$. That concludes the proof.
\end{proof}

\begin{restatelemma}{\bf \ref{lem:town_decomp}.}[Lemma 3.3 in \cite{feldmann}]

For every town $T$ in a town decomposition $\mathcal{T}$,
\begin{enumerate}[noitemsep]
\item $T$ has either 0 children or at least 2 children, and 
\item if $T$ is a town at level $i$ and has child town $T'$ at level $j$, then 
$j<i$.
\end{enumerate}
\end{restatelemma}
\begin{proof}
 We sketch the proof from \cite{feldmann}. The first property comes from the 
facts that every singleton is a town at level $0$ and that if $T'$ is a child 
town of $T$, then $T \setminus T' \ne \emptyset$. The second property is a 
consequence of the first one combined with the isolation of the child town $T'$.
\end{proof}

\section{Proof of Lemma~\ref{lem:depot_safety}}\label{sec:add_depot}

In Section~\ref{sec:prelims} we emphasize that the shortest-path covers are 
required
to be inclusion-wise minimal, and that we cannot simply add the depot to the 
shortest-path
cover at every scale. In fact, doing so actually \emph{is} safe.  This is not 
immediately
obvious, as this
modification can greatly alter the town decomposition.  However, the only risk 
in
adding a hub to the shortest-path cover is exceeding the bound on the doubling 
dimension
of $X_T$ for some town $T$.  Indeed, the only place where Feldmann et al. 
\cite{feldmann}
depend on the shortest-path covers being inclusion-wise minimal is in the proof 
of
this doubling-dimension bound. It is fairly simple to adapt their proofs
to show that adding a fixed number of vertices to a minimal shortest-path cover 
at
every levels adds at most a small factor to
the bound on the doubling dimension.

Instead of reproving their results, however, we modify the graph to give the
desired property.  We now prove the claimed properties of the 
Lemma~\ref{lem:depot_safety}
modification (restated for convenience).

\begin{restatelemma}{\bf \ref{lem:depot_safety}.}
Any graph $G = (V,E)$ with highway dimension $\hdimension$,
diameter $\Delta_G$, and designated vertex set  $S \subseteq V$ can be modified 
by
adding
$O(\hdimension^2|S|^3\log\Delta_G)$ new vertices and edges, such
that the resulting graph $G'=(V',E')$ 
\begin{itemize}
\item has highway dimension at most $\hdimension+|S|$ 
\item for all $u,v, \in V'$, $d_{G'}(u,v) \in (\frac{c} {2},\ \frac{3c} 
{4}\Delta_G]$
\item for all $u,v \in V$, $d_{G'}(u,v) = d_G(u,v)$, and 
\item for every $s \in S$, the only towns containing $s$ in the town 
decomposition
of $G'$ are the trivial towns.
\end{itemize}
\end{restatelemma}

\begin{figure}[H]     
\centering  
\includegraphics[width=.8\textwidth]{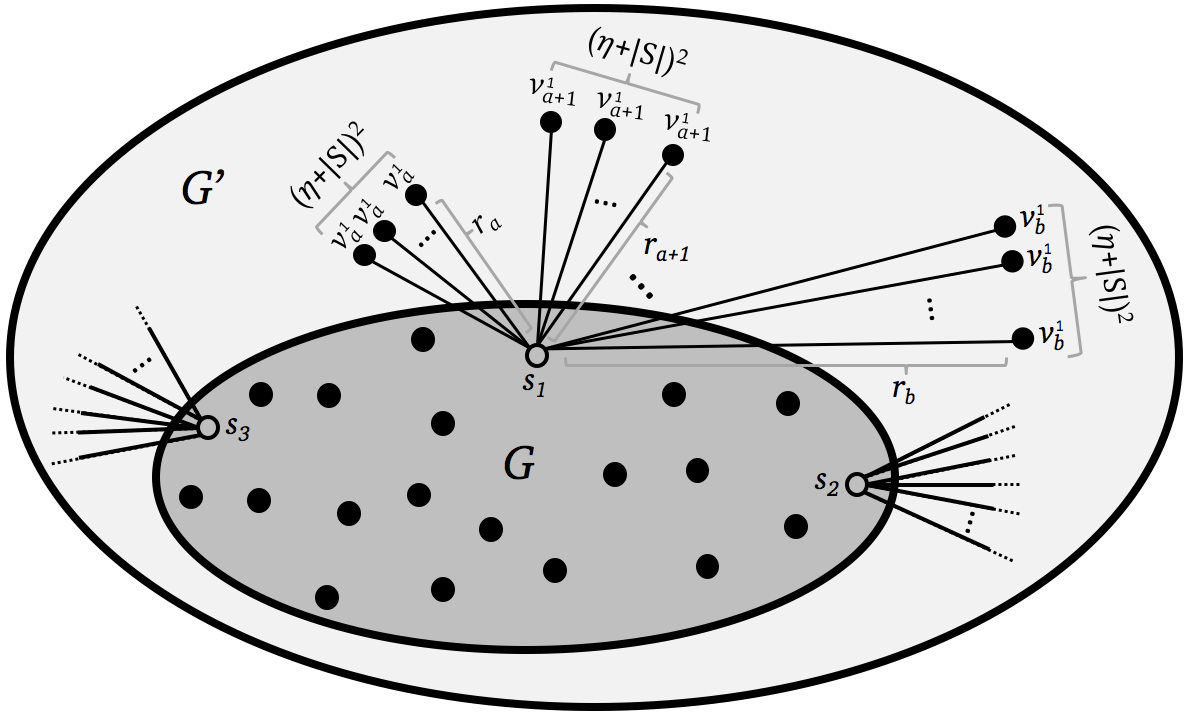}          
\caption{\label{fig:depot_mod} Here, $S = \{s_1,s_2,s_3\}$
is depicted as hollow vertices in $G$. For each $s \in S$, the modification of 
$G$
to $G'$ introduces new vertices
and edges between these new vertices and $s$. }
\end{figure}

\begin{proof}

Let $a\in \mathbb{Z}$ be the smallest integer such that $(\frac{c}{4})^a > 
\frac{c}{2}$ and
let $b \in \mathbb{Z}$ be the smallest integer such that $(\frac{c}{4})^b > 
\Delta_G$.
 We modify $G$ by adding, for each $i \in \{a,\ a+1, ...,\ b\}$ and each $s\in 
S$,
 $(\hdimension+|S|)^2$ copies of vertex $v^s_i$ and an edge $(s,v^s_i)$ of 
length
 $r_i
 = (\frac{c} {4})^i$ for each copy.  This modification adds $(\hdimension+|S|)^2 
|S|
 (b-a+1) = O(\hdimension^2|S|^3\log\Delta_G)$
vertices and edges (see Figure~\ref{fig:depot_mod}). We show that each of the 
listed
properties holds for the modified
graph $G'$.

First, this modification increases the highway
dimension by at most $|S|$, since adding $S$ as hubs covers all
newly introduced shortest paths.  

Second the smallest introduced edge has length $(\frac{c}{4})^a
> \frac{c}{2}$, and all point-to-point distances in $G$ are already assumed to 
be
greater than $\frac{c}{2}$ (see Section~\ref{sec:prelims}).  The largest 
introduced
edge has length $(\frac{c}{4})^b > \Delta_G \geq (\frac{c}{4})^{b-1}$, so the 
largest
point-to-point
distance in $G'$ is between two copies of $v^s_b$ from distinct vertices 
$s_1,s_2\in S$, namely $\Delta_{G'} \leq 2(\frac{c}{4})^b + \Delta_G < 
3(\frac{c}
{4})^b = \frac{3c}{4}(\frac{c}{4})^{b-1} \leq \frac{3c}{4}\Delta_G$.

Third, all newly added edges are only connected to vertices in $S$, so all 
point-to-point
distances
between vertices in $V$ are preserved in $G'$.

Finally, recall that the trivial towns in the town decomposition of $G'$ are the 
singleton
towns at scale $r_0 = (\frac{c}{4})^0=1$
and the topmost town at scale $r_{max} = \ceil{\log_{c/4}diam(G')}$ that 
contains
all of $G'$. By the second
property above, all point-to-point distances in $G'$ are greater than $c/2$ and, 
clearly,
at most $diam(G')$. Therefore there are no shortest paths in $G'$ in the 
intervals
$(1, c/2]$ or $(r_{max}, \frac{c}{2}r_{max}]$, so $SPC(r_0) = SPC(r_{max}) = 
\emptyset$.
 So all of $G'$ is in the trivial town at scale $r_{max}$ and each vertex $v\in 
V'$,
including each $s\in S$, is in a trivial singleton town at scale $r_0$. Consider 
some
$s \in S$. We must show that $s$ does not appear in a town at any scale $r_i
\in [r_1,r_{max})$.

We first show that for every scale $r_i = (\frac{c}{4})^i \in 
[\frac{c}{4},(\frac{c}
{4})^{b}) = [r_1,r_b)$, every locally-sparse shortest-path cover 
SPC$\big({(\frac{c}
{4})^i}\big)$
of $G'$ includes $s$, and therefore $s$ cannot be in any town at these scales. 
The
shortest path cover SPC$\big({(\frac{c} {4})^i}\big)$ must contain a hub on each
path with length in $((\frac{c}{4})^i, \frac{c}{2}(\frac{c}{4})^i]$. By the 
first
property
above, SPC$\big({(\frac{c} {4})^i}\big)$ is guaranteed to be locally 
$(\hdimension
+ |S|)\log
(\hdimension+|S|)$-sparse
\cite{abraham11}, so
$\big|B_s(c(\frac{c} {4})^i) \cap \text{SPC($(\frac{c} {4})^i$)}\big| \leq 
(\hdimension
+ |S|)\log(\hdimension+|S|) < (\hdimension+|S|)^2$. There are two cases to 
consider.

If $\frac{c}{4} \leq (\frac{c}{4})^i < (\frac{c}{4})^a$,
then $(\frac{c}{4})^a \in ((\frac{c}{4})^i, \frac{c}{2}(\frac{c}{4})^i]$, since 
$
(\frac{c}{4})^{a-1} \leq \frac{c}{2}$ implies $(\frac{c}{4})^{a} \leq 
\frac{c}{2}
(\frac{c}{4}) \leq \frac{c}{2}(\frac{c}{4})^i$.  Since each edge
$(s,v^s_a)$ has length
$(\frac{c}{4})^{a}$, SPC$\big({(\frac{c} {4})^i}\big)$ must contain either $s$ 
or all
$(\hdimension+|S|)^2$ copies of $v^s_a$. By the sparsity argument above, 
SPC$\big({(
\frac{c}
{4})^i}\big)$ must therefore contain $s$.

Otherwise, $(\frac{c}{4})^a \leq (\frac{c}{4})^i < (\frac{c}
{4})^{b}$.  Therefore, $(\frac{c}{4})^{i+1} \leq (\frac{c}{4})^b$, so there are 
$
(\hdimension+|S|)^2$ newly added edges $(s,v^s_{i+1})$ of length 
$(\frac{c}{4})^{i+1}$.
Furthermore $(\frac{c} {4})^{i+1} \in ((\frac{c}{4})^i, 
\frac{c}{2}(\frac{c}{4})^i]$,
since $\frac{c}{4} < \frac{c}{2}$.
Therefore SPC$\big({(\frac{c} {4})^i}\big)$ must contain either
$s$ or all $(\hdimension+|S|)^2$ copies of $v^s_{i+1}$. Again, by the sparsity 
argument
above, SPC$\big({(\frac{c}{4})^i}\big)$ must therefore contain $s$.

What remains to show is that $s$ is in no town at scales in $[r_b,r_{max})$.  We 
show
that in fact there are no non-trivial towns at these levels.  Assume to the 
contrary
that $T$
is a non-trivial town at scale $r_i \in [r_b,r_{max})$.  By 
Lemma~\ref{lem:towns},
 $d_{G'}(T,V'\setminus T) > r_i$.  In particular, any edges in the cut 
$\delta(T)$
 must have length greater than $r_i$.  Furthermore since $T$ is non-trivial, $T 
\neq
 V'$, and since $G'$ is connected, $\delta(T) \neq \emptyset$.  Therefore there
 is some edge in $G'$
 that has
 length greater than $r_i$ and thus greater than $r_b$.  However all edges from 
$G$
 have length at most $\Delta_G < r_b$, and all newly added edges have length
 at
 most $r_b$.  Therefore no such town exists.
 
\end{proof}

\end{document}